\PassOptionsToPackage{dvipsnames}{xcolor}

\documentclass{article}
\usepackage{bbm}
\usepackage[utf8]{inputenc} 
\usepackage[T1]{fontenc}    
\usepackage{hyperref}       
\usepackage{url}            
\usepackage{booktabs}       
\usepackage{amsfonts}       
\usepackage{nicefrac}       
\usepackage{microtype}      
\usepackage{lipsum}		
\usepackage{graphicx}
\usepackage{float}
\usepackage{natbib}
\usepackage{doi}
\usepackage{xcolor}
\usepackage[normalem]{ulem}

\usepackage{color}
\usepackage{lineno}
\usepackage{amsmath}
\usepackage{xcolor}

\usepackage{soul}
\usepackage{arxiv}
\usepackage{url}
\usepackage{bm}
\newcommand\myshade{85}
\colorlet{mylinkcolor}{RoyalBlue}
\colorlet{mycitecolor}{violet}
\colorlet{myurlcolor}{YellowOrange}

\hypersetup{
  linkcolor  = mylinkcolor!\myshade!black,
  citecolor  = mycitecolor!\myshade!black,
  urlcolor   = myurlcolor!\myshade!black,
  colorlinks = true,
}

\usepackage{times}
\usepackage{latexsym}

\usepackage[T1]{fontenc}

\usepackage[utf8]{inputenc}

\usepackage{microtype}

\usepackage{booktabs}
\usepackage{graphicx}
\usepackage{array}
\usepackage{multirow}
\usepackage{longtable}
\usepackage{CJKutf8}
\usepackage{float}
\usepackage{microtype}
\usepackage{multirow}
\usepackage{booktabs}
\usepackage{graphicx}
  
\usepackage{amssymb}
\usepackage{url}
\usepackage{hyperref}
\usepackage{amsmath}
\usepackage{caption}
\usepackage[ruled,vlined,linesnumbered]{algorithm2e}
\usepackage{subfigure}
\usepackage{makecell}
\usepackage{ulem}
\usepackage{algorithmic}
\usepackage[misc]{ifsym}
\usepackage{lipsum}
\usepackage{colortbl}

\usepackage{tikz}
\usetikzlibrary{shapes.geometric}

\usepackage{palatino} 

\usepackage{graphicx}
\usepackage{multirow}
\usepackage[most]{tcolorbox}
\usepackage{subcaption}
\usepackage{svg}

\usepackage{tabularx}
\usepackage{adjustbox}
\usepackage{amssymb} 
\usepackage{pifont} 

\usepackage{verbatim}

\usepackage[utf8]{inputenc}
\usepackage{pifont}
\usepackage{newunicodechar}

\newcommand*\colourcheck[1]{%
  \expandafter\newcommand\csname #1check\endcsname{\textcolor{#1}{\ding{52}}}%
}
\colourcheck{blue}
\colourcheck{green}
\colourcheck{red}

\newcommand*\colourmark[1]{%
  \expandafter\newcommand\csname #1mark\endcsname{\textcolor{#1}{\ding{55}}}%
}
\colourmark{blue}
\colourmark{green}
\colourmark{red}

\title{CodeEditorBench: Evaluating Code Editing Capability of Large Language Models}

\newcommand\blfootnote[1]{%
\begingroup
\renewcommand\thefootnote{}\footnote{#1}%
\addtocounter{footnote}{-1}%
\endgroup
}

\author{ 
    \textbf{Jiawei Guo\textsuperscript{1}$^{\ast}$},\space
    \textbf{Ziming Li\textsuperscript{3}$^{\ast}$},\space
    \textbf{Xueling Liu\textsuperscript{1}$^{\ast}$},\space
    \textbf{Kaijing Ma\textsuperscript{5}$^{\ast}$},\space
    \\
    \textbf{Tianyu Zheng\textsuperscript{1}},\space
    \textbf{Zhouliang Yu}\textsuperscript{3},\space
    \textbf{Ding Pan\textsuperscript{3}},\space
    \textbf{Yizhi Li\textsuperscript{4}},\space 
    \\
    \textbf{Ruibo Liu\textsuperscript{1}},\space
    \textbf{Yue Wang\textsuperscript{1}},\space
    \textbf{Shuyue Guo\textsuperscript{1}},\space
    \textbf{Xingwei Qu\textsuperscript{3,4}},\space 
    \\
    \textbf{Xiang Yue\textsuperscript{1}},\space
    \textbf{Ge Zhang\textsuperscript{1,2,6}$^{\dag}$},\space
    \textbf{Wenhu Chen\textsuperscript{1,2,6}$^{\dag}$},\space
    \textbf{Jie Fu\textsuperscript{3}$^{\dag}$},\space
    \\
{\small 
\textsuperscript{1} Multimodal Art Projection Research Community;
}
{\small 
\textsuperscript{2} University of Waterloo;
} \\
{\small 
\textsuperscript{3} HKUST;
} 
{\small 
\textsuperscript{4} University of Manchester;
}
{\small 
\textsuperscript{5} Tongji University;
}
{\small 
\textsuperscript{6} Vector Institute
}
}

\begin{document}
\maketitle

\vspace{-7ex}
\begin{center}
     \url{https://codeeditorbench.github.io}
\end{center}
\vspace{3ex}

\begin{abstract}

Large Language Models (LLMs) for code are rapidly evolving, with code editing emerging as a critical capability. We introduce CodeEditorBench, an evaluation framework designed to rigorously assess the performance of LLMs in code editing tasks, including debugging, translating, polishing, and requirement switching. Unlike existing benchmarks focusing solely on code generation, CodeEditorBench emphasizes real-world scenarios and practical aspects of software development. We curate diverse coding challenges and scenarios from five sources, covering various programming languages, complexity levels, and editing tasks. Evaluation of 19 LLMs reveals that closed-source models (particularly Gemini-Ultra and GPT-4), outperform open-source models in CodeEditorBench, highlighting differences in model performance based on problem types and prompt sensitivities. 
CodeEditorBench aims to catalyze advancements in LLMs by providing a robust platform for assessing code editing capabilities. We will release all prompts and datasets to enable the community to expand the dataset and benchmark emerging LLMs. By introducing CodeEditorBench, we contribute to the advancement of LLMs in code editing and provide a valuable resource for researchers and practitioners. 

\end{abstract}

\blfootnote{$^{\ast}$ These authors contributed equally.}
\blfootnote{$^{\dag}$ Corresponding Authors.}

\section{Introduction}

\begin{figure}[h]
\begin{center}
\includegraphics[width=\linewidth]{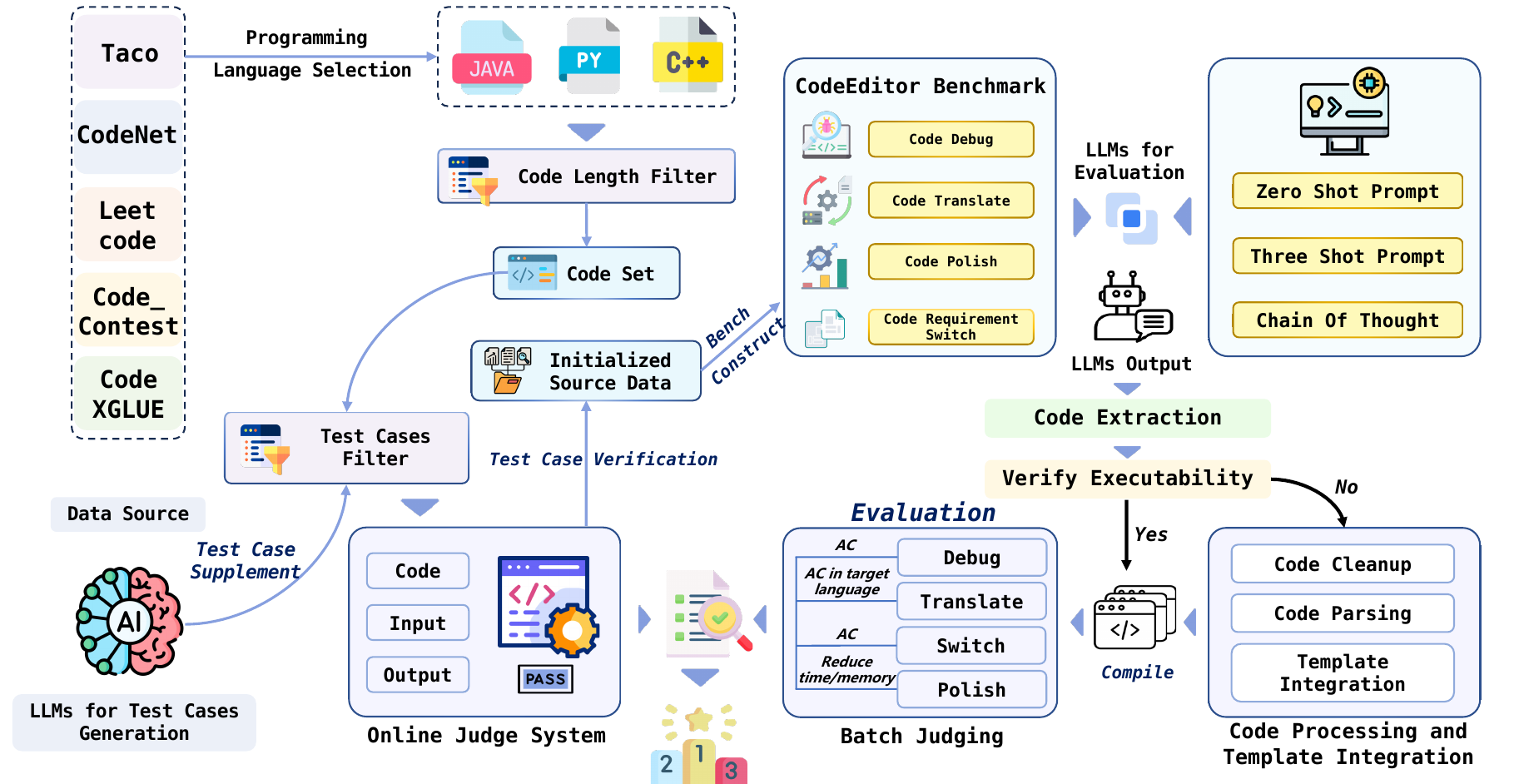}
\end{center}
\caption{Overview of \textit{CodeEditorBench}.
CodeEditorBench evaluates programming languages by selecting initial data from five sources and filtering based on code length. It enriches the dataset with Large Language Model-generated test cases, which, along with all code, are verified by an Online Judge System (OJ). The benchmark is developed for four problem types using specific methodologies, described in Section~ \ref{method}. Assessment of 19 LLMs involves crafting prompts for zero-shot, three-shot, and chain of thought settings. Outputs are filtered and integrated with templates for compilation. The OJ's batch judging determines the LLMs' scores, ensuring a rigorous evaluation process.
}

\label{fig:technical_route}
\vspace{-0.2cm}
\end{figure}

Recent advancements in LLMs~\citep{touvron2023llama, achiam2023gpt} underscore the importance of their coding capabilities, extending beyond mere programming assistance~\citep{tian2023chatgpt,nijkamp2023codegen2}
to encompass various tool-using applications~\citep{qin2023toolllm, cai2023large}.
Specifically,  code LLMs~\citep{rozière2024code, guo2024deepseekcoder,zheng2024opencodeinterpreter} are deployed across various tasks, such as code repair~\citep{olausson2023demystifying}, code optimization~\citep{shypula2023learning}. 
Despite their increasing adoption as programming aids, existing evaluation methods~\citep{chen2021evaluating, austin2021program_mbpp} primarily focus on code generation, neglecting the crucial aspect of code editing in software development.
To bridge such a significant gap in evaluation, we advocate for developing a new benchmark to comprehensively assess the code editing abilities of these LLMs. 

Coding encompasses a wide range of skills, with code editing playing a pivotal role in software development. 
This constitutes a significant portion of the development process, incorporating a variety of sub-tasks such as code optimization, refactoring, and bug fixing, each presenting unique challenges. 
Code LLMs' efficacy, adaptability, and functionality are paramount for enhancing development efficiency and code quality and facilitating team collaboration.
Code editing is distinctly different from code completion. 
Therefore, datasets frequently employed in code completion studies, such as HumanEval~\citep{chen2021evaluating} and MBPP~\citep{austin2021program}, are inadequate for assessing code editing capabilities.

Motivated by the shortcomings, we introduce \textbf{CodeEditorBench}, a pioneering evaluation framework designed to assess the performance of LLMs in editing code rigorously, where the overview is described in Figure \ref{fig:technical_route}.
The categorization of code editing problems helps to understand and evaluate the performance of code LLMs systematically. 
Based on the SDLC definitions\footnote{\url{https://en.wikipedia.org/wiki/Software_development_process}}, we evaluate code LLMs in four scenarios: Code Debug, Code Translate, Code Polish, and Code Requirement Switch, which are categorized to reflect the most common and critical types of tasks in the code editing process.
See Appendix~\ref{classification_basis} for classification basis.
\begin{itemize}
    \item \textbf{Code Debug}: Debugging is the process of locating and fixing errors in code. 
    \item \textbf{Code Translate}: Translating means converting code from one programming language to another. 
    \item \textbf{Code Polish}: Polishing refers to optimizing code without changing its functionality.
    \item \textbf{Code Requirement Switch}: Requirement switching is adapting code to new or modified requirements. 
\end{itemize}

Additionally, we make significant efforts to manually construct test cases for each problem in various programming scenarios to precisely check the editing correctness. We further established an online evaluation system to facilitate easy evaluation of a broad set of code LLMs.
Consequently, we compiled a dataset containing 7,961 code editing tasks, each with an average of 44 test cases (a minimum of 8 and a maximum of 446).
Inspired by LiveCodeBench~\citep{jain2024livecodebench}, we implemented a timestamp-based filtering process to consider the data pollution phenomenon. This process led to a refined dataset called CodeEditorBench\_Plus. The original dataset was thereafter designated as CodeEditorBench\_Primary. In Section~\ref{subsec:Data Construction}, we detailed the dataset construction process.

To complement the introduction of a comprehensive benchmark, we aim to delineate the current array of available models through a code editing leaderboard. We assess 6 base models and 13 models that undergo instruction tuning across four distinct scenarios, utilizing the same experimental framework, and employ two evaluative approaches: zero-shot and three-shot. 

\begin{figure*}[htbp]
    \centering
    \begin{minipage}{0.45\textwidth}
        \centering
        \includegraphics[width=\textwidth]{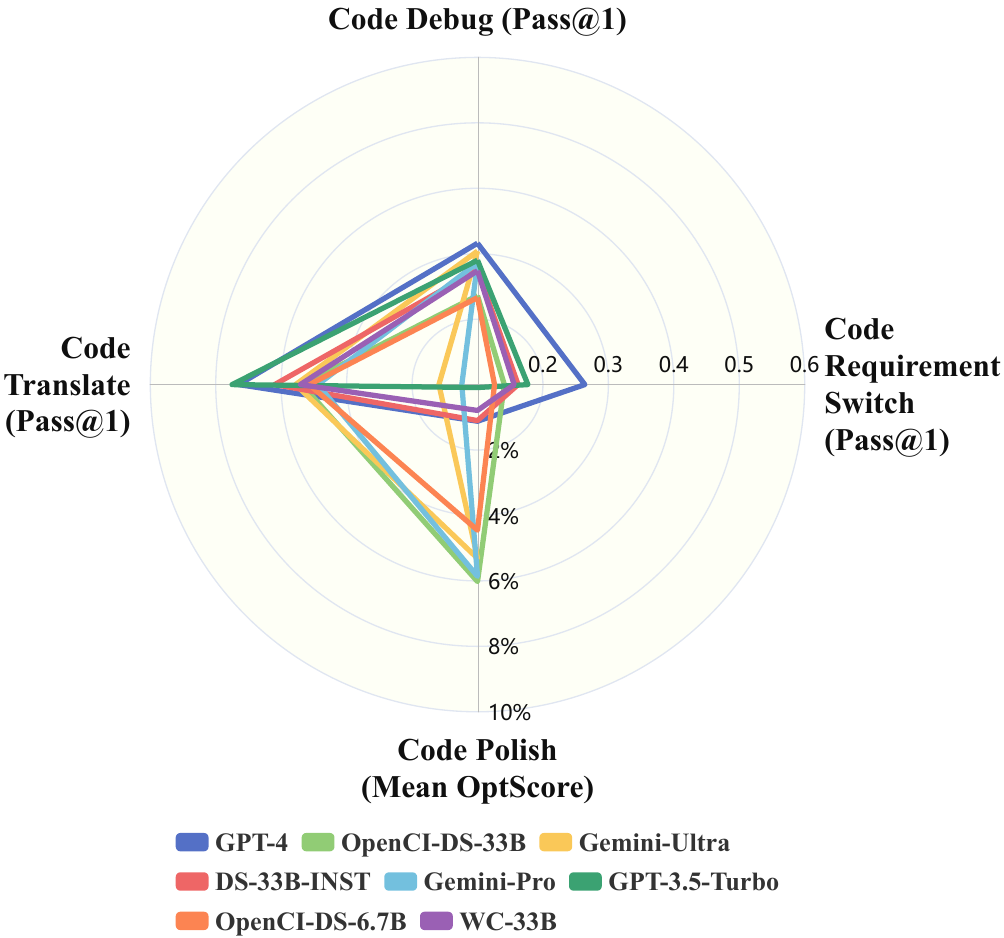}
    \end{minipage}
    \hfill
    \begin{minipage}{0.49\textwidth}
        \centering
        \includegraphics[width=\textwidth]{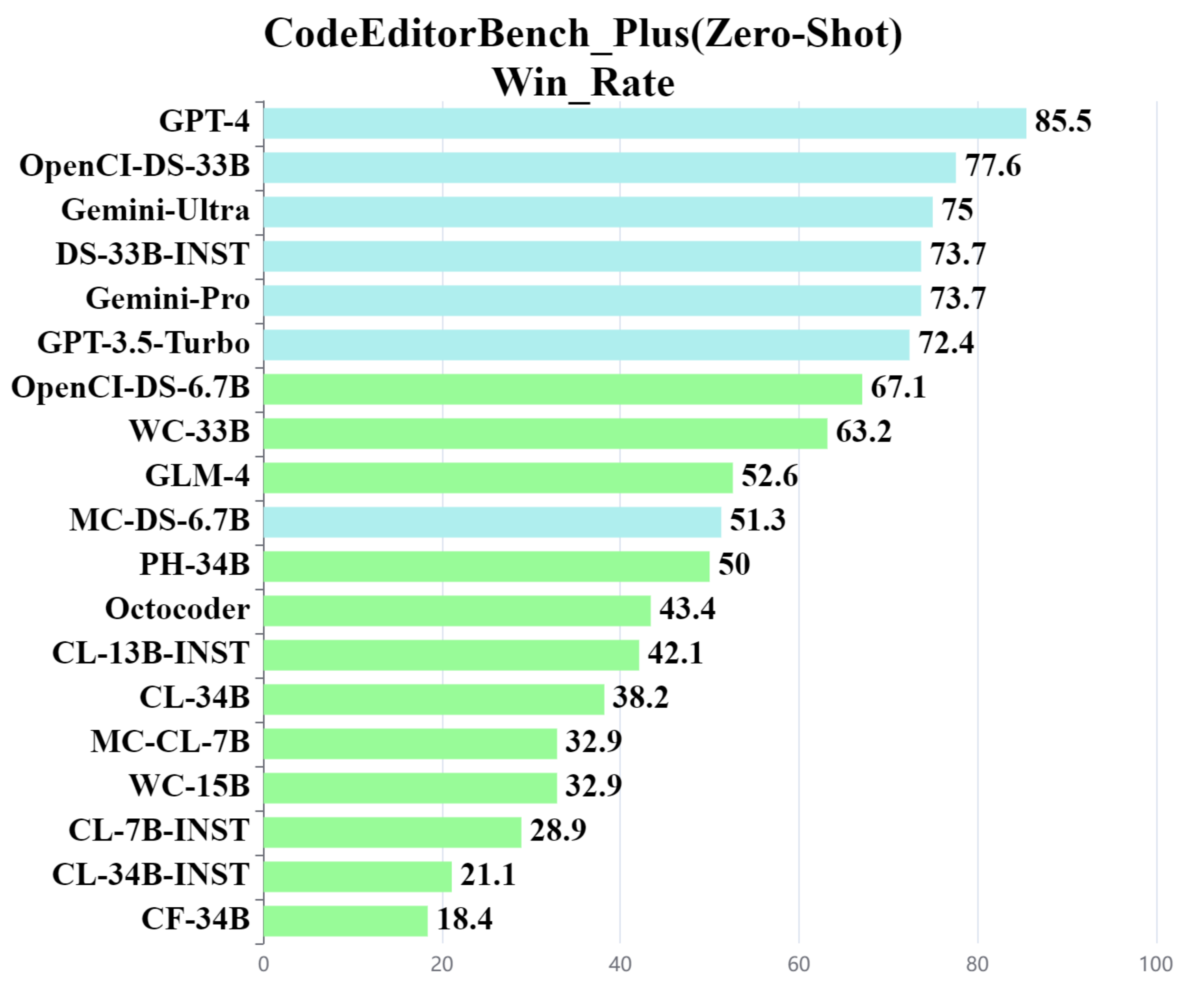}
    \end{minipage}
    \hfill
    \vspace{5pt}
    \caption{Based on Section \ref{win_rate}. \textbf{Left}. We propose evaluating LLMs across four scenarios capturing various code editing capabilities, namely code debug, code translate, code polish, and code requirement switch. The figure depicts various model performances across the four scenarios available in CodeEditorBench\_Plus in a radial plot – highlighting how relative differences across models change across the scenarios.
    \textbf{Right}. Performance of open-source and closed-source models on CodeEditorBench\_Plus in zero-shot evaluated through win\_rate.
    For a comprehensive explanation of the abbreviation, refer to Section \ref{subsec:evaluatedmodels}.}
    \label{fig:mix_zero}
    \vspace{-0.2cm}
\end{figure*}

With CodeEditorBench, we aim to:
\begin{itemize}
    \item Provide a unified framework for assessment, including tools for visualization, training, and additional analyses. We will also make all the data involved in the evaluations publicly available to foster further examination of LLM characteristics. Furthermore, we plan to incorporate more evaluation metrics in the future.

    \item Map the current landscape of LLMs. Among the publicly accessible base models, OpenCI-DS-33B emerges as the most potent, followed by OpenCI-DS-6.7B and DS-33B-INST. Models without open access, like Gemini, GPT, and GLM, typically surpass openly available ones. However, instruction-tuned models with more than 30 billion parameters, namely OpenCI-DS-33B and DS-33B-INST, bridge this performance divide. We present detailed results in Figure~\ref{fig:mix_zero}, Figure~\ref{fig:mix_few} and Table~\ref{tab:All_Results}.

    \item Highlight the models’ limitations in code polishing and code rewriting as required. While GPT-4 shows commendable performance in three out of four areas, its capabilities in code polishing are notably inadequate. Similarly, Gemini Ultra falls short in tasks involving code requirement alterations.
\end{itemize}

\section{Related Work}

\textbf{Code LLMs} \quad The field witnesses significant growth in developing code LLMs to address the challenges in code understanding and generation. 
This trend starts with the introduction of Codex~\citep{chen2021evaluating} by OpenAI, followed by the emergence of many influential models including CodeGen~\citep{nijkamp2023codegen}, CodeT5~\citep{wang2021codet5, wang2023codet5}, and InCoder~\citep{fried2023incoder}.
Recent popular open-source models like CodeLLaMa~\citep{rozière2024code}, DeepSeek Coder~\citep{guo2024deepseekcoder}, and StarCoder~\citep{li2023starcoder, lozhkov2024starcoder} represent the forefront of this field. 
They demonstrate excellent abilities in various code understanding and generation tasks by extensively pre-training from scratch on massive code datasets. 
Additionally, these base models undergo another phase of instruction tuning~\citep{zheng2024opencodeinterpreter, wei2023magicoder, luo2023wizardcoder, Royzen2023Phind, mftcoder2023, muennighoff2024octopack}, empowering them with better instruction-following capability, leading to significant performance improvements in solving various code-related tasks.

\textbf{Code Benchmark} \quad Many benchmarks are proposed to compare and evaluate code LLMs.
However, these primarily focus on natural language and code generation.
HumanEval~\citep{chen2021evaluating} is one of the pioneering and most widely used benchmarks for LLM-based code synthesis, consisting of 164 pairs of Python function signature with docstring and the associated test-cases for correctness checking. 
Another Python-focused dataset, MBPP~\citep{austin2021program}, is created by crowd-sourcing participants to write in summation 974 programming problems, each of which is comprised of the problem statement (i.e., docstring), the function signature, as well as three test-cases. 
Beyond Python, there are other benchmarks targeting additional languages such as Spider (SQL)~\citep{yu2018spider}, HumanEval-X~\citep{zheng2023codegeex} (C++, Javascript and Go), HumanEvalPack~\citep{muennighoff2024octopack}(Python, JavaScript, Java, Go, C++ and Rust), CodeContests~\citep{li2022competition} (C++ and Java) and MultiPL-E~\citep{cassano2022multipl} (extending HumanEval and MBPP to 18 programming languages). 
Competitive programming benchmarks include LeetCode-Hard Gym~\citep{olausson2023demystifying}, which evaluates code generation in multiple languages using LeetCode's server and the OpenAI gym framework. DebugBench~\citep{tian2024debugbench}  advances LLM evaluation by focusing on error correction across diverse programming challenges, from syntax to logical mistakes.
EditEval~\citep{hu2023instructcoder} assesses LLMs' ability to understand and execute code editing instructions, measuring how accurately models can modify code based on human-written instructions. 

Relatively little work addresses the objective of code editing.
Previous works either focus on a subset of code editing tasks or do not give a reasonable division of code edit tasks.
To fill this gap, we introduce CodeEditorBench, a pioneering evaluation framework designed to assess the performance of LLMs in editing code.

\section{Method}

\label{method}
\subsection{Problem Definition}
For any question $q_i$, in order to test whether the code $c_i$ generated by LLM is correct, a reasonable approach is to construct a set $R_i$ - a collection of test cases containing test case input-output pairs ($x_i$,$y_i$), each $x_i$ being a program input and $y_i$ being a corresponding desired output.
Let $a_c$($x$) = $y$ denote a program a, based on a code script $c$, that maps an input x to an output y.
We define the answer code $c_i$ to be correct if and only if there is not any input-output pair in $R_i$ such that $a_{c_i}(x_i) \neq y_i$
\begin{itemize}

\item an ideal debugger D that rectifies any buggy code from $c$ to $c*$ should satisfy that D($c$) = $c*$  s.t.$\forall$($x_i$, $y_i$) $\in R_i$, $a_c*$($x_i$) = $y_i$. Debugging can be regarded as the converting process of debugger D.

\item an ideal translater T that translates any code from language a to language b should satisfy that T($c$) = $c*$ s.t.  $\forall$($x_i$,$y_i$) $\in R_i$, $a_c$($x_i$) = $y_i$ and $a_c*$($x_i$) = $y_i$. Translating can be regarded as the converting process of translater T.

\item an ideal optimizer P that make any code more effective in either time complexity or space complexity with the guarantee of accuracy should satisfy that P($c$) = $c*$  s.t.  $\forall$($x_i$,$y_i$) $\in R_i$, $a_c$($x_i$) = $y_i$,$a_c*$($x_i$) = $y_i$ and $avg_{time}(c*) \leq avg_{time}(c)$ or $avg_{memory}(c*) \leq avg_{memory}(c)$.
Polishing can be regarded as the converting process of optimizer P

\item an ideal requirement switcher S that switches a similar code that satisfies question $a$ to a target code that satisfies question $b$ based on the sample inputs and outputs of $a$ and $b$ should satisfy that S($c$) = $c*$ s.t. $\forall$($x_i$,$y_i$) $\in R_b$, $a_c*$($x_i$) = $y_i$.Switching can be regarded as the converting process of switcher S.
\end{itemize}

\begin{figure}[h]
\begin{center}
\includegraphics[width=\linewidth]{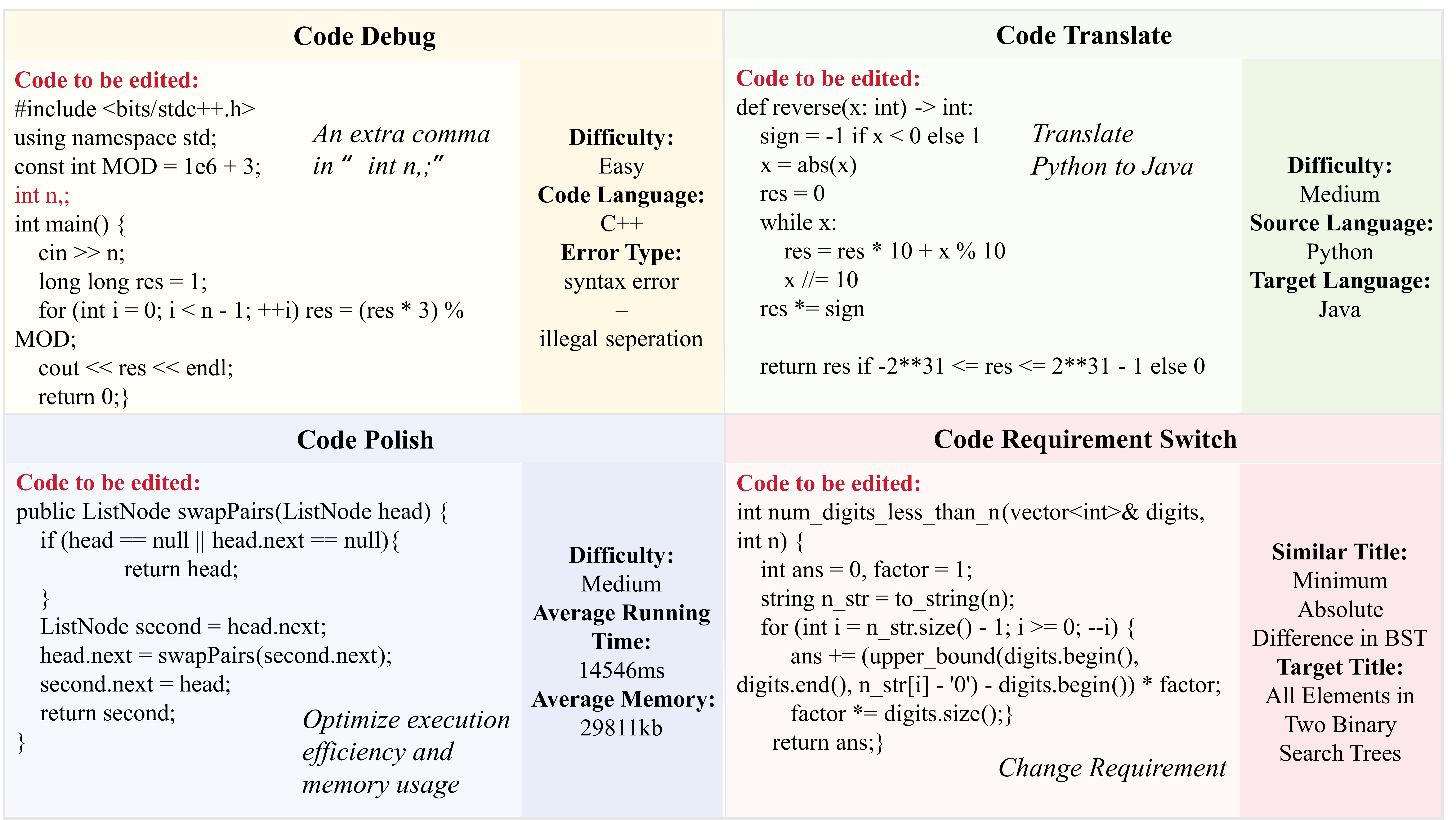}
\end{center}
\caption{Data Samples of CodeEditorBench}
\label{fig:setting_sample}
\vspace{-0.4cm}
\end{figure}

\subsection{Data Construction}
\label{subsec:Data Construction}

In our meticulously crafted compilation, we gather a wide-ranging assortment of coding challenges sourced from five sources: namely leetcode\footnote{\url{https://leetcode.com}}, code\_contests~\citep{li2022competition}, CodeXGLUE~\citep{DBLP:journals/corr/abs-2102-04664}, codeNet~\citep{puri2021codenet} and Taco~\citep{li2023taco}. To align with the constraints imposed by the maximum token limit of advanced LLMs, we applied a stringent selection criterion, setting aside any question codes that surpassed 800 lines or exceeded a 1000-token threshold. This approach ensures the integrity of the code generated by LLMs.

Our collection spans extensive data structures— from trees, stacks, and queues to arrays, hash tables, and pointers. 
It also covers a broad spectrum of algorithms, including but not limited to dynamic programming, various sorting techniques, depth-first search (DFS), breadth-first search (BFS), greedy algorithms, and recursion. 
This comprehensive compilation aims to facilitate an all-encompassing exploration and understanding of coding principles, catering to various computational problems and scenarios.

\subsubsection{Construction Method}

\textbf{Code Debug} \quad Inspired by the DebugBench~\citep{tian2024debugbench},
we employ the insertion of a basic error to construct the data with one error and utilized a similar methodology to generate data sets with two, three, and four errors, as detailed in the basic error types in Table~\ref{code-error-basic-type}. The precise prompts used for inserting errors are detailed in Appendix~\ref{insert_error}.

\textbf{Code Translate \& Code Polish} \quad  Based on a straightforward rationale, the challenge LLMs encounter in understanding code typically increases as the complexity of the code itself rises. Consequently, we stratify the dataset according to code complexity, adhering to a selection ratio of 3:4:1.

\textbf{Code Requirement Switch} \quad  Initially, we categorize the data into two groups. The first is designated as 'strong relation', which represents the similar questions provided by Leetcode under a certain question.
These questions exhibit clear human feedback and represent the most tightly interconnected subjects. Conversely, the second category, 'weak relation', which is constructed by us. The methodology for its construction is outlined as follows:
\begin{itemize}
\item Collect the labels of each question.
\item Cluster the questions based on the number of tags they possess, with our clustering criterion being the presence of four or more identical tags.
\item Given that tags only partially convey the essence of the questions, we employ Bert~\citep{devlin2018bert} to assess the semantic similarity between the descriptions of two questions within each category. Our set threshold for similarity is 0.92.
\end{itemize}

Despite the majority of the dataset being synthetically generated by us, we acknowledge the inherent risk of data pollution within the dataset. 
To address this concern, we implement a timestamp-based filtering process. 
This novel strategy enables us to methodically examine the dataset, thereby identifying and excluding outdated information, which significantly improves the overall quality of the dataset. 
Consequently, from CodeEditorBench\_Primary, we develop CodeEditorBench\_Plus.

\subsubsection{Data Post-processing}

\textbf{Code template construction} To address issues with LeetCode code submissions lacking header files, starter code, and other necessary components, we develop specialized code templates, which can be found in Appendix \ref{code_template}. 
Focusing on C++ as an illustrative case, our contribution comprises two main components: \texttt{FastIO.h} and \texttt{LeetcodeIO.h}.

The FastIO component introduces an efficient mechanism for input and output operations, either with standard input/output streams or files. This design aims to surpass the performance constraints of traditional C++ I/O methods. 
The LeetcodeIO component builds upon FastIO's capabilities, emulating the input and output behavior characteristic of the LeetCode online coding platform. 
This enables local testing of solutions that mirrors the LeetCode environment.

Both components prioritize simplicity in integration and usage, necessitating only the inclusion of the respective header files and slight modifications to the existing code framework. 
Additionally, they provide extra features like logging and customizable file path options for I/O tasks, thus facilitating an improved debugging and testing workflow.

\textbf{Test cases generation} During the benchmark construction, we encounter several issues: notably, some topics are deficient in test cases, failing to meet the minimum requirement of eight cases. 
Additionally, the comprehensiveness of some test cases is compromised due to the absence of boundary tests, thereby falling short of the standards for rigorous evaluation. 
To address these shortcomings, we leverage GPT-4, GLM-4, and Qwen-72B-chat for test case generation. 
Mindful of the potential for illusion, we opt to select only the inputs from the generated test cases. 
The corresponding outputs are then determined by executing the problem-solving code via OJ, which we establish to assess the correctness of code generated by LLMs.
The specific prompt for generating test cases can be found in Figure~\ref{generate_test_case} and Figure~\ref{generate_boundary}.
\subsection{Data Analysis}

Figure~\ref{fig:setting_sample} presents a selection of exemplars from the CodeEditorBench, delineating the spectrum of code editing tasks, including Code Debugging, Code Translating, Code Polishing, and Code Requirement Switching. 
Each dataset entry shares similar attributes such as title, difficulty, public and private test inputs, and outputs, as well as \texttt{code\_language}. 
Unique to the Code Debugging tasks is an 'error\_type' attribute, denoting the specific issue present in the code. 
Translating tasks are distinguished by \texttt{source\_lang} and \texttt{target\_lang} attributes, indicating the original and target programming languages. 
Code Polishing tasks include \texttt{avg\_running\_time} and \texttt{avg\_memory} attributes, quantifying the average execution time and memory usage, respectively. 
Finally, Code Requirement Switching tasks feature a distinct set of attributes including \texttt{public\_similar\_tests}, \texttt{public\_target\_tests} and \texttt{private\_target\_tests} inputs and outputs, which correspond to the initial and target test cases of the coding problem. And the same is true for \texttt{similar\_title} and \texttt{target\_title}.

The CodeEditorBench\_Primary, as illustrated in Figure~\ref{fig:four_code_primary}, and CodeEditorBench\_Plus, as shown in Figure~\ref{fig:four_code_plus}, establish an evaluation framework that mirrors the complexities inherent in real-world software development scenarios. This framework is meticulously designed to gauge LLMs' code editing capabilities, presenting a richer and more nuanced set of challenges than conventional code generation benchmarks. The datasets are extensive, categorizing tasks along several dimensions: programming languages (C++, Java, Python), number of errors (one, two, three, four), difficulty levels (easy, medium, hard), language transitions (e.g., C++ to Java), and relation strength (strong, weak).

\section{Experiments}
\label{sec:experiments}

\subsection{Experimental Setup}
\textbf{Decoding Strategy.} Our dataset is significantly larger than those used in prior benchmarks, necessitating adjustments in our evaluation approach. 
For closed-source models, given the considerable expense associated with API calls, we opt for greedy decoding when generating code, employing the pass@1~\citep{kulal2019spoc, chen2021evaluating} metric to assess the code's pass rate. 
We also apply a greedy decoding strategy to open-source models to facilitate a fair comparison between them. 

\textbf{Hyperparameter Settings.} Our dataset was filtered to exclude code exceeding 800 lines or 1024 tokens. 
In our benchmark's four scenarios, the generated code is typically similar in length to the source code, leading us to set the count of maximum new tokens to 1024 for closed-source models. 
For open-source models, we set the maximum new tokens' count to 2048. 
We utilize vLLM~\citep{kwon2023efficient} to accelerate the output process for open-source models. 

The same experimental setup is used across all four scenarios.

\subsection{Evaluated Models}
\label{subsec:evaluatedmodels}
We select 19 of the most popular current code LLMs from existing leaderboards evalplus~\citep{liu2023code}, bigcode~\citep{bigcode-evaluation-harness}, encompassing both open-source and closed-source models, with sizes ranging from 6.7B to 34B, including base models and instruction-tuning models. 
The open-source models include the GPT series (GPT-3.5-Turbo, GPT-4), the Gemini series (Gemini-Pro, Gemini-Ultra), and GLM-4. 
Closed-source models comprise the CodeLlama series (CL-34B-Base, CL-\{7, 13, 34\}B-INST), the DeepSeek series (DS-33B-INST), and the outstanding instruction-finetuned models, the WizardCoder series (WC-15B based on StarCoder, WC-33B based on DS-33B-Base), OctoCoder based on StarCoder, CF-34B based on CL-34B-Python, PH-34B based on CL-34B-Base, the Magicoder series (MC-DS-6.7B based on DS-6.7B-Base, MC-CL-7B based on CL-7B) and the OpenCodeInterpreter series(OpenCI-DS-6.7B based on DS-6.7B-Base, OpenCI-DS-33B based on DS-33B-Base). 
Appendix~\ref{app:Evaluated Models} presents more detailed information regarding our evaluated models.

\subsection{Prompt Engineering}
\textbf{Prompt Setting.} We implement various prompting techniques to evaluate our model, including zero-shot prompting and few-shot prompting methods. 
In selecting examples for few-shot prompts, a clustering approach is utilized to pick three fixed examples from the dataset of each scenario. 

\textbf{Prompt Format.} In constructing prompts for open-source models, we adhere to the formats provided in their official examples. 
We search for a suitable prompt format in the HuggingFace model card, Github repository, and formal publications or technical reports. 
The variance in prompts across different open-source models primarily lies in a few special identifiers, such as for the CodeLlama-INST series of models, we add \text{$<$s$>$[INST] and [/INST]} at the beginning and end of the instruction, respectively. 
We consistently use the \textit{Instruction-Question-Answer} format to construct prompts for closed-source models.

\section{Result Analysis}
\label{sec:result_analysis}
\subsection{Evaluation Online System} 
To comprehensively assess LLM's code editing performance across the four scenarios, we construct OJ based on the hustoj~\citep{Haobin2012DesignAI}.
The system processes LLM-generated code to ensure adherence to operational requirements, tailors a set of focused test problems, and passes criteria for each scenario, facilitating a comprehensive evaluation.

\begin{table}[!tb]
\centering
\scalebox{0.85}{
\begin{tabular}{lcccccccc}\toprule
Model &Size &Open &Debug &Translate &Switch &Polish &Win Rate \\\midrule
\multicolumn{8}{c}{Zero-shot} \\
\midrule
GPT-4                                             & - & \redmark & \textbf{0.316}(0.493) & 0.465(\textbf{0.503})    & \textbf{0.264}  & 1.12\%(1.33\%)    & \textbf{0.855}(\textbf{0.868})     \\
OpenCI-DS-33B              & 33B & \greencheck & 0.236(0.429) & 0.368(0.428)    & 0.141  & \textbf{6.02\%}(6.49\%)    & 0.776(0.816)     \\
Gemini-Ultra                                     & - & \redmark & 0.304(0.459) & 0.378(0.278)    & 0.041  & 5.31\%(3.77\%)    & 0.750(0.579)     \\
DS-33B-INST            & 33B & \greencheck & 0.275(0.487) & 0.410(0.451)    & 0.162  & 1.10\%(1.14\%)    & 0.737(0.757)     \\
Gemini-Pro                                        & - & \redmark & 0.286(0.423) & 0.344(0.344)    & 0.076  & 5.86\%(\textbf{6.65\%})    & 0.737(0.711)     \\
GPT-3.5-Turbo                                       & - & \redmark & 0.290(\textbf{0.494}) & \textbf{0.475}(0.480)    & 0.177  & 0.09\%(0.84\%)    & 0.724(0.776)     \\
OpenCI-DS-6.7B            & 6.7B & \greencheck & 0.233(0.402) & 0.357(0.384)    & 0.126  & 4.45\%(4.28\%)    & 0.671(0.697)     \\
WC-33B                   & 33B & \greencheck & 0.274(0.487) & 0.371(0.438)    & 0.156  & 0.79\%(0.90\%)    & 0.632(0.704)     \\
GLM-4                                   & - & \redmark  & 0.220(0.271) & 0.278(0.365)    & 0.085  & 5.17\%(6.46\%)    & 0.526(0.592)     \\
MC-DS-6.7B                   & 6.7B & \greencheck & 0.242(0.406) & 0.343(0.401)    & 0.130  & 0.21\%(1.99\%)    & 0.513(0.697)     \\
PH-34B                 & 34B & \greencheck & 0.230(0.369) & 0.279(0.331)    & 0.074  & 2.84\%(1.78\%)    & 0.500(0.539)     \\
Octocoder                                 & 15.5B & \greencheck & 0.042(0.145) & 0.392(0.223)    & 0.030  & 1.39\%(2.70\%)    & 0.434(0.289)     \\
CL-13B-INST              & 13B & \greencheck & 0.176(0.368) & 0.333(0.275)    & 0.021  & 2.31\%(1.82\%)    & 0.421(0.368)     \\
CL-34B                        & 34B & \greencheck & 0.163(0.250) & 0.310(0.240)    & 0.052  & 1.10\%(0.84\%)    & 0.382(0.171)     \\
MC-CL-7B                      & 7B & \greencheck & 0.174(0.317) & 0.272(0.276)    & 0.039  & 1.31\%(1 31\%)    & 0.329(0.342)     \\
WC-15B                   & 15B & \greencheck & 0.159(0.354) & 0.309(0.278)    & 0.067  & 0.91\%(0.96\%)    & 0.329(0.408)     \\
CL-7b-INST               & 7B & \greencheck & 0.155(0.336) & 0.289(0.231)    & 0.017  & 1.47\%(1.17\%)    & 0.289(0.250)     \\
CL-34B-INST              & 34B & \greencheck & 0.131(0.250) & 0.287(0.240)    & 0.027  & 1.02\%(0.84\%)    & 0.211(0.171)     \\
CF-34B                  & 34B & \greencheck & 0.166(0.223) & 0.218(0.177)    & 0.028  & 0.33\%(0.70\%)    & 0.184(0.105)     \\
\midrule
\multicolumn{8}{c}{Few-shot} \\
\midrule
Gemini-Ultra                  & - & \redmark & 0.286(0.448) & 0.443(0.307)    & 0.152  & 5.62\%(4.55\%)    & \textbf{0.855}(0.632)     \\
GPT-4                                  & - & \redmark & \textbf{0.345}(\textbf{0.523}) & \textbf{0.517}(\textbf{0.514})    & \textbf{0.303}  & 1.13\%(1.14\%)    & 0.816(\textbf{0.882})     \\
OpenCI-DS-6.7B & 6.7B & \greencheck & 0.233(0.440) & 0.372(0.399)    & 0.165  & \textbf{6.47\%}(\textbf{8.59\%})    & 0.770(0.750)     \\
OpenCI-DS-33B   & 33B & \greencheck & 0.230(0.463) & 0.371(0.437)    & 0.229  & 5.75\%(4.82\%)    & 0.763(0.803)     \\
DS-33B-INST & 33B & \greencheck & 0.272(0.489) & 0.417(0.465)    & 0.235  & 1.18\%(0.93\%)    & 0.737(0.763)     \\
GPT-3.5-Turbo                                 & - & \redmark & 0.270(0.511) & 0.364(0.431)    & 0.201  & 1.54\%(1.70\%)    & 0.684(0.803)     \\
Gemini-Pro                    & - & \redmark & 0.229(0.386) & 0.392(0.356)    & 0.139  & 5.23\%(5.64\%)    & 0.671(0.645)     \\
WC-33B        & 33B & \greencheck & 0.279(0.515) & 0.362(0.447)    & 0.243  & 0.65\%(0.63\%)    & 0.645(0.711)     \\
MC-DS-6.7B        & 6.7B & \greencheck & 0.262(0.478) & 0.321(0.381)    & 0.192  & 1.44\%(0.89\%)    & 0.605(0.632)     \\
GLM-4                                  & - & \redmark & 0.233(0.341) & 0.299(0.360)    & 0.100  & 5.30\%(6.41\%)    & 0.572(0.592)     \\
CL-34B             & 34B & \greencheck & 0.133(0.367) & 0.307(0.252)    & 0.113  & 1.75\%(1.11\%)    & 0.474(0.447)     \\
PH-34B      & 34B & \greencheck & 0.239(0.468) & 0.275(0.326)    & 0.092  & 1.20\%(0.75\%)    & 0.421(0.461)     \\
CL-13B-INST   & 13B & \greencheck & 0.160(0.330) & 0.327(0.284)    & 0.028  & 1.75\%(1.25\%)    & 0.414(0.322)     \\
MC-CL-7B           & 7B & \greencheck & 0.157(0.355) & 0.245(0.230)    & 0.075  & 1.70\%(1.18\%)    & 0.329(0.382)     \\
WC-15B        & 15B & \greencheck & 0.114(0.332) & 0.271(0.224)    & 0.099  & 1.65\%(1.11\%)    & 0.322(0.329)     \\
CF-34B       & 34B & \greencheck & 0.166(0.262) & 0.240(0.158)    & 0.050  & 1.61\%(1.39\%)    & 0.289(0.250)     \\
Octocoder                      & 15.5B & \greencheck & 0.050(0.263) & 0.290(0.206)    & 0.054  & 1.09\%(0.85\%)    & 0.211(0.184)     \\
CL-7B-INST    & 7B & \greencheck & 0.167(0.362) & 0.271(0.224)    & 0.028  & 1.00\%(0.71\%)    & 0.211(0.204)     \\
CL-34B-INST   & 34B & \greencheck & 0.143(0.330) & 0.303(0.264)    & 0.032  & 0.32\%(0.67\%)    & 0.211(0.211)     \\
\bottomrule
\end{tabular}
}
\vspace{10pt}
\caption{Evaluating LLMs on CodeEditorBench. All results of models are generated by greedy decoding. Code Debug, Code Translate and Code Requirement Switch are evaluated with pass@1, while Code Polish is evaluated with Mean OptScore. 
Values outside parentheses denote Plus results and inside denote Primary results. For the Switch class, Primary and Plus results are identical, and only one score is displayed.}
\label{tab:All_Results}
\vspace{-0.3cm}
\end{table}

\textbf{Pass Criteria.}
For Code Debug, Code Translate, and Code Requirement Switch, we verify whether the code passes all test cases within the time and memory constraints of our OJ. 
For Code Translate, we specifically run the code in the target language environment to ensure an accurate evaluation of translation performance. 
To meet the pass criteria for Code Polish, the code must pass all test cases and demonstrate improved efficiency by reducing execution time or memory usage relative to the original version.

For more detailed information on the pass criteria and evaluation configuration, please refer to Appendix ~\ref{app:pass} and Appendix ~\ref{app:eval_configure} respectively.


\subsection{Performance Metrics}
\label{win_rate}
We evaluate the 19 models described in Section~\ref{subsec:evaluatedmodels} using a zero-shot and few-shot approach on CodeEditorBench. 
For Code Debug, Code Translate, and Code Requirement Switch, we employ pass@1 evaluation criteria. For Code Polish, we use the Mean OptScore as a ranking metric. 
We measure the average runtime $\bar{T}$ and average memory usage $\bar{M}$ over 20 executions of the original code. 
For each polish problem, we conduct two measurements of the model-generated code and calculate the average values $\bar{T}_{\text{avg}}$ and $\bar{M}_{\text{avg}}$. 
For each model-generated code pass all test cases, 
if $\bar{T} > \bar{T}_{\text{avg}}$ or $\bar{M} > \bar{M}_{\text{avg}}$, the code is considered to pass, and the score is calculated as:
\[OptScoreTime = \frac{(\bar{T} - \bar{T}_{\text{avg}})}{\bar{T}} \]
\[OptScoreMem = \frac{(\bar{M} - \bar{M}_{\text{avg}})}{\bar{M}} \]
\[ OptScore = \max\left[\frac{(OptScoreTime + OptScoreMem)}{2}, 0\right] \]

Otherwise, the score is set to 0. We calculate the average OptScore across questions and obtain the mean OptScore.

Finally, we calculate each model's rank based on its performance on CodeEditorBench PLUS for each problem category. 
Inspired by ~\citep{bigcode-evaluation-harness}, we utilize the win rate to evaluate the model's overall performance across various types of problems.
We compute the win rate using $1 - (rank - 1) / num\_models$ for each problem category, and average them across all categories as the win rate. 
The results are summarized in Table~\ref{tab:All_Results}.

\subsection{Model Performance Comparison}

\begin{figure}[h]
\begin{center}
\includegraphics[width=0.9\linewidth]{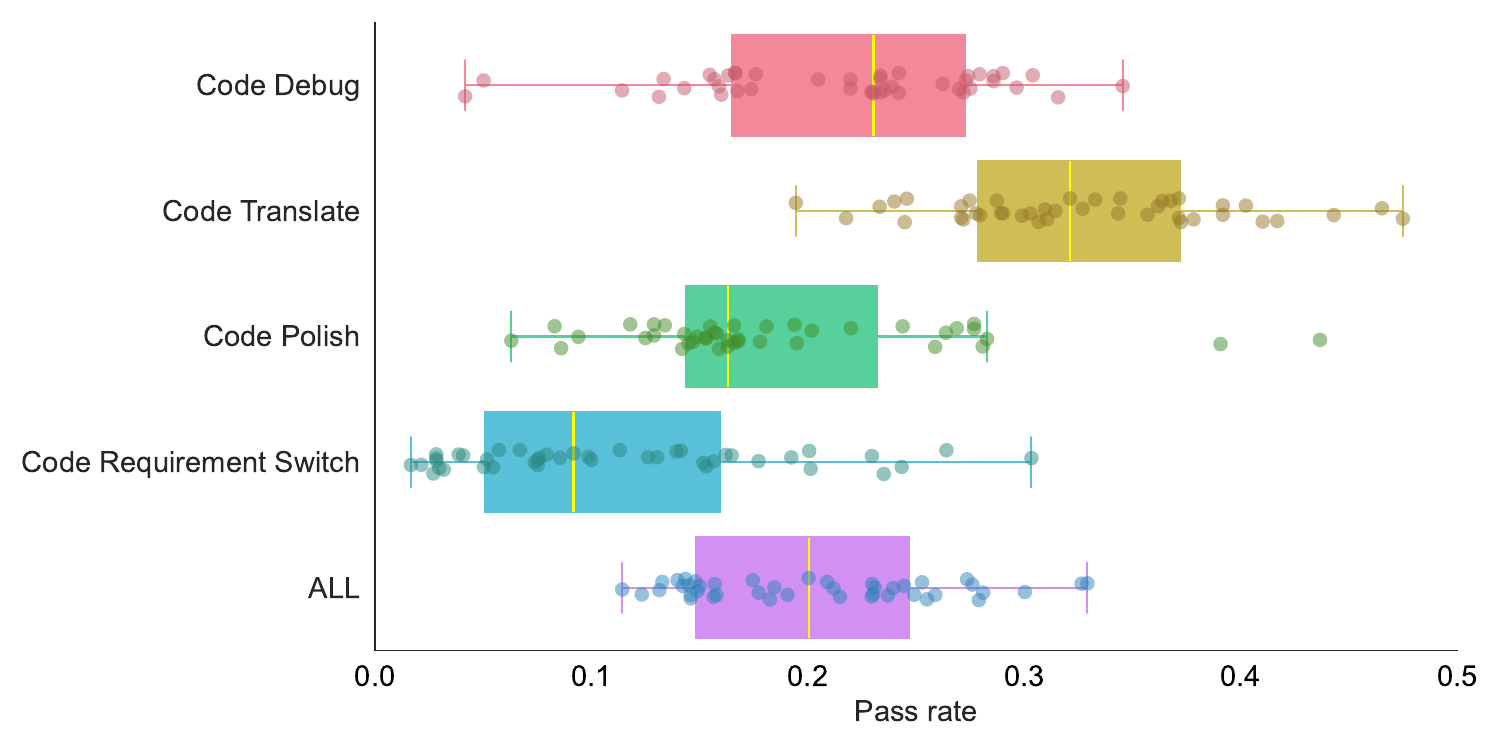}
\end{center}
\caption{Pass rate distribution
of models on CodeEditorBench\_Plus}
\label{fig:pass_rate_boxplot}
\end{figure}

Analysis of Table~\ref{tab:All_Results} reveals that, in Debug and Translate data types, some LLMs exhibit a significant difference in pass@1 between Primary and Plus datasets, with values exceeding 0.2. 
This discrepancy suggests the potential for data leakage within the Primary dataset.
Therefore, we focus on analyzing the performance of LLMs on the Plus dataset.

\textbf{Situation}. The comparative analysis of model efficacy on the Plus dataset, delineated in Table~\ref{tab:All_Results}, underscores that closed-source LLMs predominantly surpass their open-source analogs. Notably, Gemini-Ultra and GPT-4 present considerable performance superiority, with GPT-4 outperforming in zero-shot scenarios and Gemini-Ultra in few-shot scenarios. Specifically, GPT-4 excels in Debug, Translate, and Switch categories, highlighting the variation in model performance across diverse problem types. Conversely, while Gemini-Ultra and Gemini-Pro are adept in Polish tasks, they falter in Switch tasks, falling behind select closed-source models. In contrast, GPT-4 and GPT-3.5 demonstrate proficiency in areas where Gemini models underperform.
As for open-source models, OpenCI-DS-33B emerges as a frontrunner in zero-shot scenarios, even eclipsing Gemini-Ultra, but mirrors the Gemini series' deficiencies in Switch tasks. OpenCI-DS-6.7B outperforms its open-source contemporaries and surpasses Gemini-Pro and GPT-3.5 in few-shot scenarios, showcasing its capability to excel beyond several closed-source models in particular scenarios. Remarkably, despite its smaller architecture, OpenCI-DS-6.7B's performance transcends much larger models, including those of 34B scale.

\textbf{Pass Rate Distribution}.
The pass rates exhibit significant variation across different problem types, as illustrated in Table~\ref{tab: judgement_type} and Figure~\ref{fig:pass_rate_boxplot}. 
The PLUS dataset identifies Switch problems as the most challenging, with a mere 11.18\% pass rate. Debug and Translate problems exhibit pass rates of approximately 20\% and 30\%, respectively. For Polish problems, even with the correct original code provided, only 37.47\% of the solutions meet all testing criteria. Additionally, only a limited 19.23\% of the dataset passes all tests and demonstrates superior average runtime or memory efficiency relative to the original code. It is also significant to note that a notable fraction of solutions simply replicate the original code with no alterations.


\begin{table}[!htp]\centering
\scalebox{1.0}{
\begin{tabular}{lccccc}\toprule
&Pass &Wrong Answer &Runtime Error &Compile Error &Other \\\midrule
Debug &21.57\% &53.41\% &13.40\% &7.51\% &4.11\% \\
Polish &19.23\% &53.15\% &5.46\% &3.01\% &19.15\% \\
Switch &11.18\% &64.74\% &10.94\% &8.09\% &5.05\% \\
Translate &33.15\% &45.67\% &3.69\% &6.06\% &11.43\% \\
ALL &20.34\% &55.26\% &8.53\% &6.34\% &9.53\% \\
\bottomrule
\end{tabular}
}
    \vspace{5pt}
\caption{Judgment results across different problem types in CodeEditorBench\_Plus}
\label{tab: judgement_type}
\end{table}

\textbf{Reasons for Not Passing}. 
We analyze the aggregated solutions from all models on CodeEditorBench\_Plus, as detailed in Table~\ref{tab: judgement_type}, and discovered that only 20.34\% of solutions successfully solve the problem, with a significant 55.26\% failing due to incorrect answers. Other prevalent causes of failure include compilation and runtime errors, while instances of timeouts or exceeding memory limits are comparatively rare. Specifically, 6.34\% of the dataset experiences compilation errors, a phenomenon that may partly stem from post-processing losses incurred during the extraction of code blocks from solutions that include textual explanations. Models producing poorly formatted output, such as OctoCoder, are notably more susceptible to compilation errors. Interestingly, Polish problems demonstrate the lowest frequencies of both runtime and compilation errors, likely attributable to the minimal alterations made to the original code by the models. Conversely, Translate problems are characterized by a lower rate of incorrect answers (45.67\%) relative to other problem types, yet suffer the highest rate of timeout errors (10.21\%).


In summary, our analysis underlines the challenges posed by CodeEditorBench and highlights areas for further research in modern software development.

\section{Conclusion}

In this study, we introduce CodeEditorBench, a pioneering benchmark created to evaluate Large Language Models (LLMs) in code editing tasks. 
CodeEditorBench is envisioned as a dynamic and scalable framework that will be periodically updated to incorporate new problems, scenarios, and models. 
Our findings indicate that closed-source models, particularly Gemini-Ultra and GPT-4, outperform open-source counterparts in CodeEditorBench\_Plus, exhibiting superior performance across various problem-solving scenarios. These models display exceptional proficiency in specific areas, including Code Polishing and Code Requirement Switching.

The analysis also underscores the variability in model performance based on problem category and scenario, revealing trends in model sensitivity to prompt formulation and highlighting instances where smaller models surpass their larger counterparts in efficiency. 
Through establishing a holistic evaluation platform, CodeEditorBench aims to foster advancements in LLMs for code editing and serve as a valuable resource for researchers and practitioners.

\section{Ethics Statement}
In conducting our research on the CodeEditorBench for LLMs in code editing tasks, we prioritize ethical considerations to ensure fairness, inclusivity, and transparency. 
Our benchmark, crafted from diverse sources, aims to represent a broad spectrum of coding challenges, languages, and tasks, mitigating biases and reflecting real-world scenarios. 
We objectively evaluate open-source and closed-source LLMs to highlight their strengths and weaknesses without bias, fostering a competitive yet fair landscape for advancements in LLM capabilities.
We are committed to open-sourcing our resources, promoting accessibility, and encouraging community engagement to refine and expand upon our findings. 
This approach promotes the democratization of coding and advances in LLM technology ethically and responsibly. 
Moreover, we carefully design our benchmark to avoid privacy, security, and bias issues, aligning with our goal of contributing positively to the field and society at large. 
Our concise ethical statement reflects our dedication to conducting our research with integrity, aiming for LLMs' broad and beneficial impact on software development.

\section{Limitations}

While our study on the CodeEditorBench introduces a novel and rigorous framework for assessing the code editing capabilities of LLMs, several limitations accompany our research. These limitations are integral to understanding our benchmark's scope, applicability, and areas for future improvement.
\textbf{Model Coverage}: The evaluation of 19 LLMs may not fully represent the extensive diversity of models available, indicating a need for a more inclusive approach in subsequent studies.
\textbf{Bias in Task Selection}: Despite efforts to increase diversity, our array of coding challenges may continue to exhibit a preference for specific languages or tasks, potentially compromising the benchmark's impartiality.
\textbf{Evaluation Metrics}: The utilized metrics may not comprehensively encompass the intricacies of code editing tasks, pointing towards a necessity for more refined assessment techniques.
\textbf{Real-world Relevance}: The benchmark simulation may not fully capture the complexity of real-world software development projects, highlighting a gap in applicability.
\textbf{Dynamic LLM Landscape}: The rapid advancement in Large Language Model (LLM) technologies may render our findings obsolete, necessitating continuous updates to the benchmark. Addressing these limitations is crucial for refining the benchmark and its enhanced utility in evaluating LLMs for code editing.



\bibliographystyle{main}
\bibliography{main}

\

\clearpage
\appendix
\onecolumn

\section{Construct Dataset}

\subsection{Classification Basis}
\label{classification_basis}

\begin{figure}[h]
\begin{center}
\includegraphics[width=0.4\linewidth]{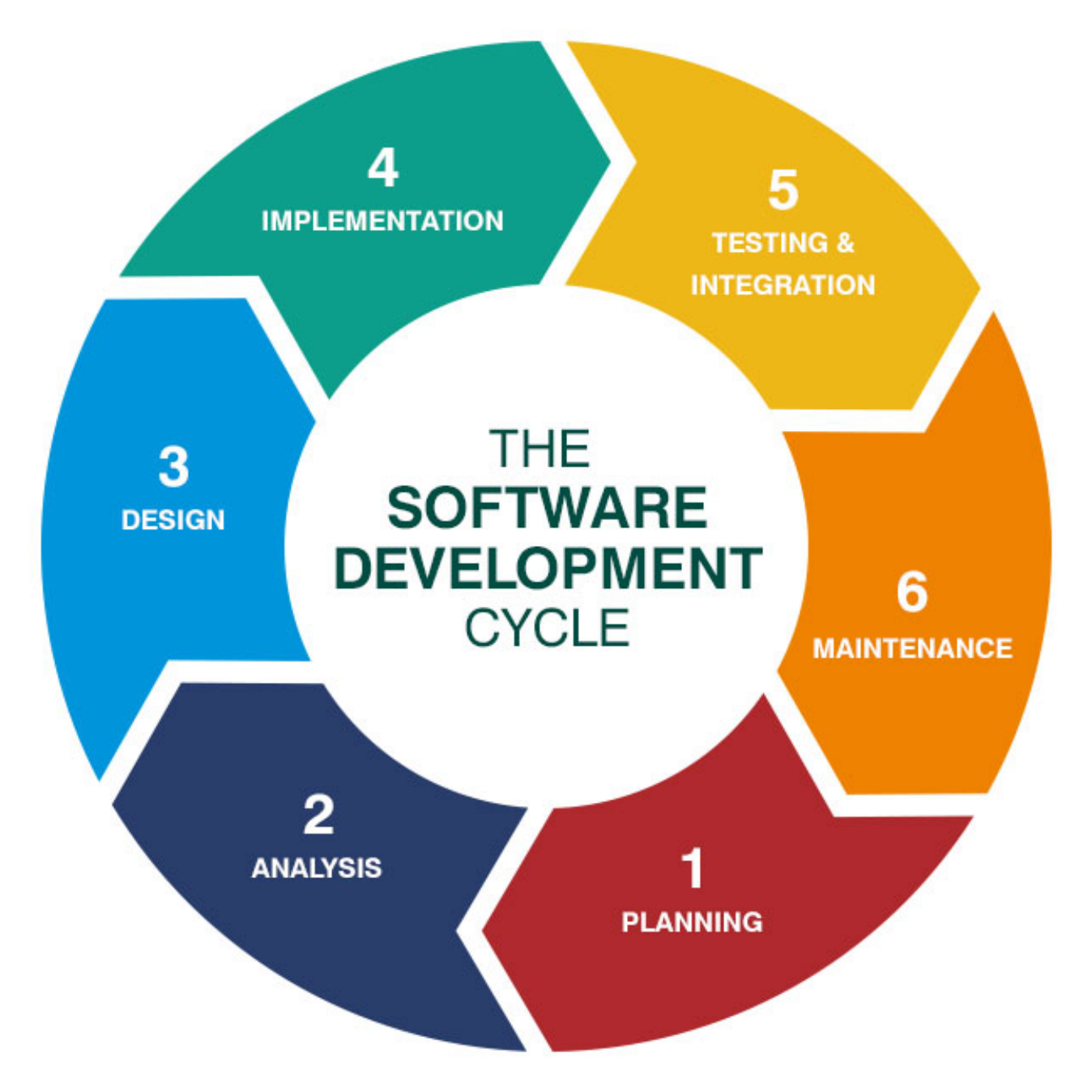}
\end{center}
\caption{Software Development LifeCycle(Source: \url{https://bigwater.consulting/2019/04/08/software-development-life-cycle-sdlc/})}
\label{fig:SDLC}
\end{figure}

The software development lifecycle (SDLC) is the cost-effective and time-efficient process that development teams use to design and build high-quality software. 
The goal of SDLC is to minimize project risks through forward planning so that software meets customer expectations during production and beyond. 
This methodology outlines a series of steps that divide the software development process into tasks you can assign, complete, and measure.

Categorizing code editing tasks based on SDLC in Figure~\ref{fig:SDLC} provides a way to understand and organize the different scenarios of code editing from the perspective of the overall flow of a software project, which typically includes phases such as planning, analysis, design, implementation, testing \& integration, and maintenance. 
Here is how to explain and categorize debug, translate, polish, and requirement switch:
\begin{enumerate}
\item \textbf{Planning.} 

Code Requirement Switch: The planning phase is the period during which the objectives, scope, schedule, and resources of the project are defined. 
During this phase, initial changes or adjustments in requirements may be identified, and the concept of requirements switching is initially developed here to meet project goals or user expectations.

\item \textbf{Analysis.} 

Code Requirement Switch: In the Requirements Analysis phase, the requirements of the project are analyzed and defined in detail. 
In this process, the requirements may be further adjusted or refined, in order to adapt to these changes, the code requirements switch in this phase becomes particularly important.

\item \textbf{Design.} 

Code Translate: The design phase is responsible for translating requirements into system architecture and detailed design. 
In this phase, the code may need to be translated to fit the design requirements, such as migrating certain components from one technology stack to another, or adapting to different platforms and frameworks.

\item \textbf{Implementation.} 

Code Polish: The implementation phase mainly involves coding. 
Code optimization and polishing are especially important in this phase to ensure the maintainability and performance of the software through refactoring, improving code structure and code quality.

Code Translate: In addition to code optimization, the implementation phase may also involve code translation, especially in multi-language programming environments or when existing code needs to be adapted to a new framework.

\item \textbf{Testing \& Integration.} 

Code Debug: The goal of the Testing \& Integration phase is to ensure the quality of the software and to find and fix any defects. 
Code debugging is a core activity in this phase to identify and resolve errors in the code to ensure that the software works as expected.

\item \textbf{Maintenance.} 

All Categories: The Maintenance phase covers all the activities that take place after the software is deployed, including fixing defects, updating the software to accommodate new requirements or changes in the environment, improving performance, and so on. In this phase:
\begin{itemize}
\item Code Debug continues to play a role in dealing with user feedback and defects found in the software.
\item Code Translate may involve code migration or rewriting efforts for compatibility or technology upgrades.
\item Code Polish focuses on improving the quality and performance of code through refactoring and optimization.
\item Code Requirement Switch reflects the need to adjust functionality and performance at any point in the software lifecycle in response to changing business requirements or user feedback.
\end{itemize}
\end{enumerate}
Through the SDLC-based categorization, we can see that 'debug', 'translate', 'polish', and 'requirement switch' are not only different aspects of code editing, but they also reflect the key tasks and challenges faced at each stage of the software development process. 
This basis for categorization emphasizes the fact that software development is a continuous, iterative process in which the activities in each phase are interdependent and work together to drive the success of a software project. 

This basis for categorization emphasizes the fact that software development is a continuous, iterative process in which the activities in each phase are interdependent and work together to drive the success of a software project.

\subsection{Insert Error}

\begin{table*}
\centering
\resizebox{\textwidth}{!}{
\renewcommand{\arraystretch}{1.2} 
\begin{tabular}{@{}lc@{}}
\hline
\textbf{Error Name} & \multicolumn{1}{c}{\bf Definition} \\ 
\midrule
misused ==/=        & Operator misuse: equality (==) vs. assignment (=)                \\ 
missing colons       & Omit colons in control structures and function definitions         \\
unclosed parentheses & Unclosed parentheses cause syntax errors                            \\
illegal separation   & Syntax errors due to improper separator usage         \\
illegal indentation  & Incorrect indentation violates syntax rules (only for Python)       \\ 
unclosed string      & Unclosed string literals: mismatched quotation marks         \\
illegal comment      & Incorrect comment syntax or placement \\
faulty indexing      & Incorrect indexing in collections leads to runtime errors                  \\
undefined objects    & Reference to undefined object: missing definition or import                 \\
undefined methods    & Calling non-existent method on object/class                \\
illegal keywords     & Reserved words misused in programming                                \\
condition error      & Logical errors in control structure conditions   \\
operation error      & Arithmetic errors, like division by zero \\
variable error       &Variable misuse errors: uninitialized variables                 \\
\hline
\end{tabular}}
\caption{\label{code-error-basic-type}
Basic Error Types
}
\end{table*}

\label{insert_error}

PS: There is an blemish in DebugBench. 
Illegal separation is a basic error defined in the debugbench.But the error exists only in java and c++ cases, not in python.
However,this error also exists in Python in reality. 
For example, 'print(a, b)' and 'print(a; b)'.So we fill that tiny gap.

\begin{figure}[h]
\begin{center}
\includegraphics[width=0.5\linewidth]{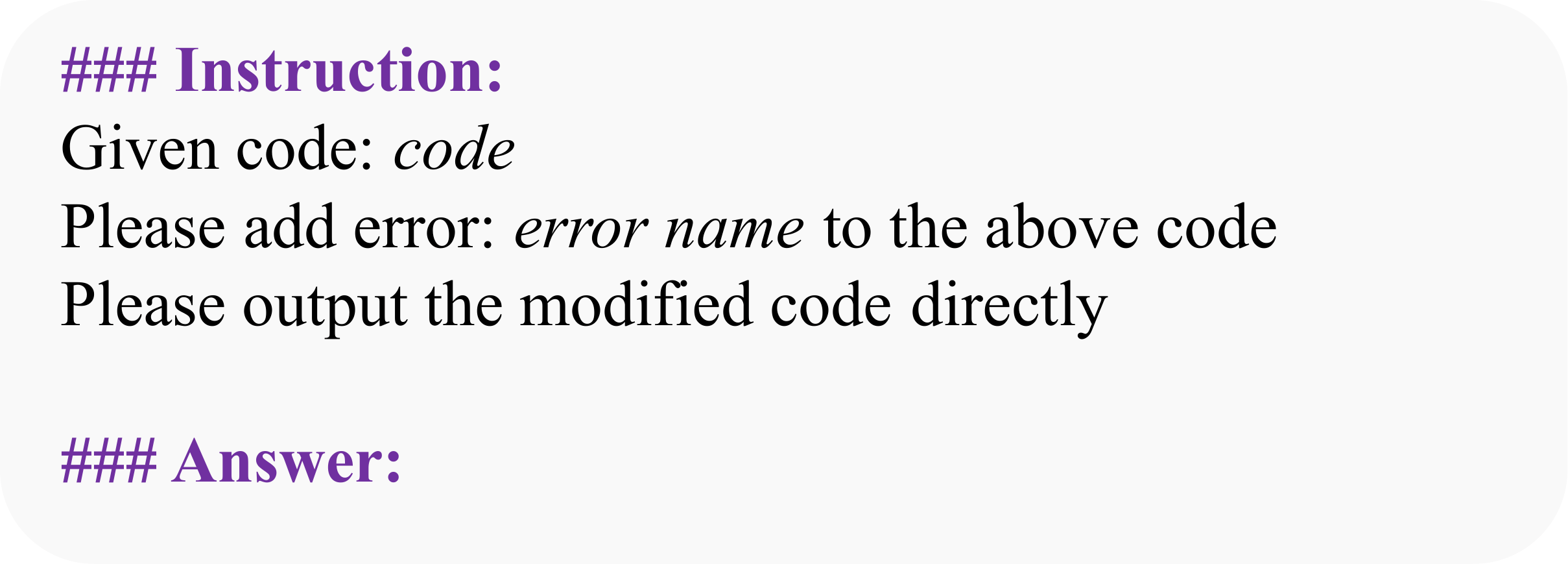}
\end{center}
\caption{Prompt for inserting basic error}
\end{figure}

\begin{figure}[h]
\begin{center}
\includegraphics[width=\linewidth]{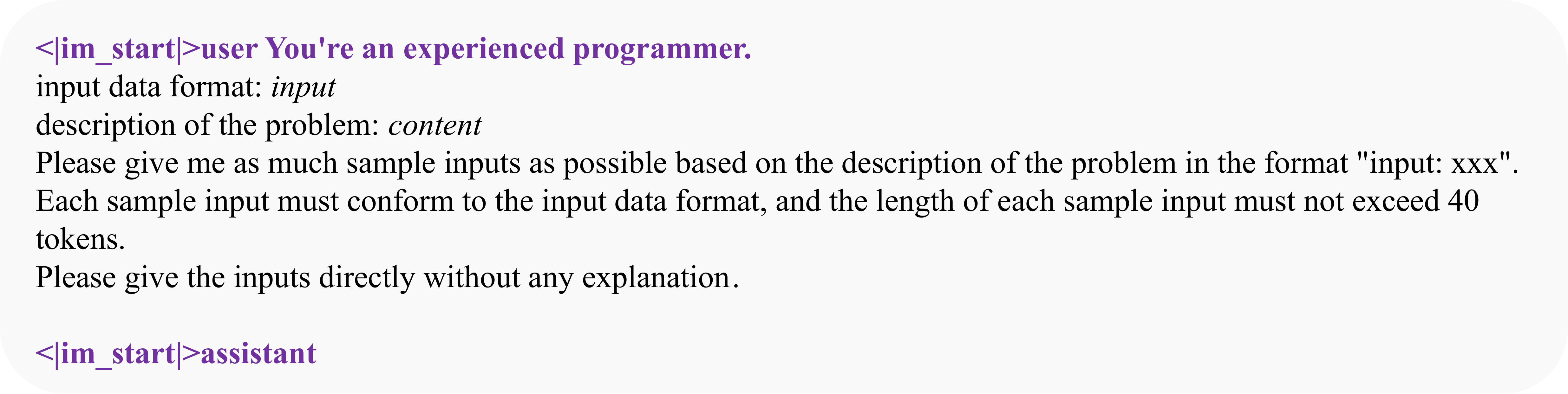}
\end{center}
\caption{Prompt for generating test cases}
\label{generate_test_case}
\end{figure}

\begin{figure}[h]
\begin{center}
\includegraphics[width=\linewidth]{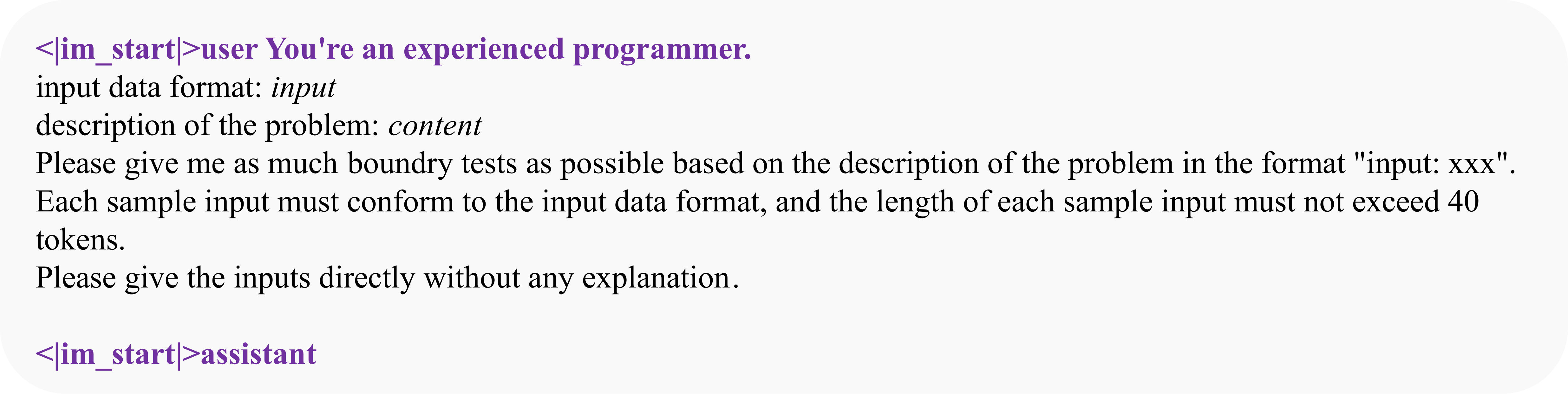}
\end{center}
\caption{Prompt for generating boundary test cases}
\label{generate_boundary}
\end{figure}

\subsection{Data Analysis}

\begin{figure}[h]
\begin{center}
\includegraphics[width=\linewidth]{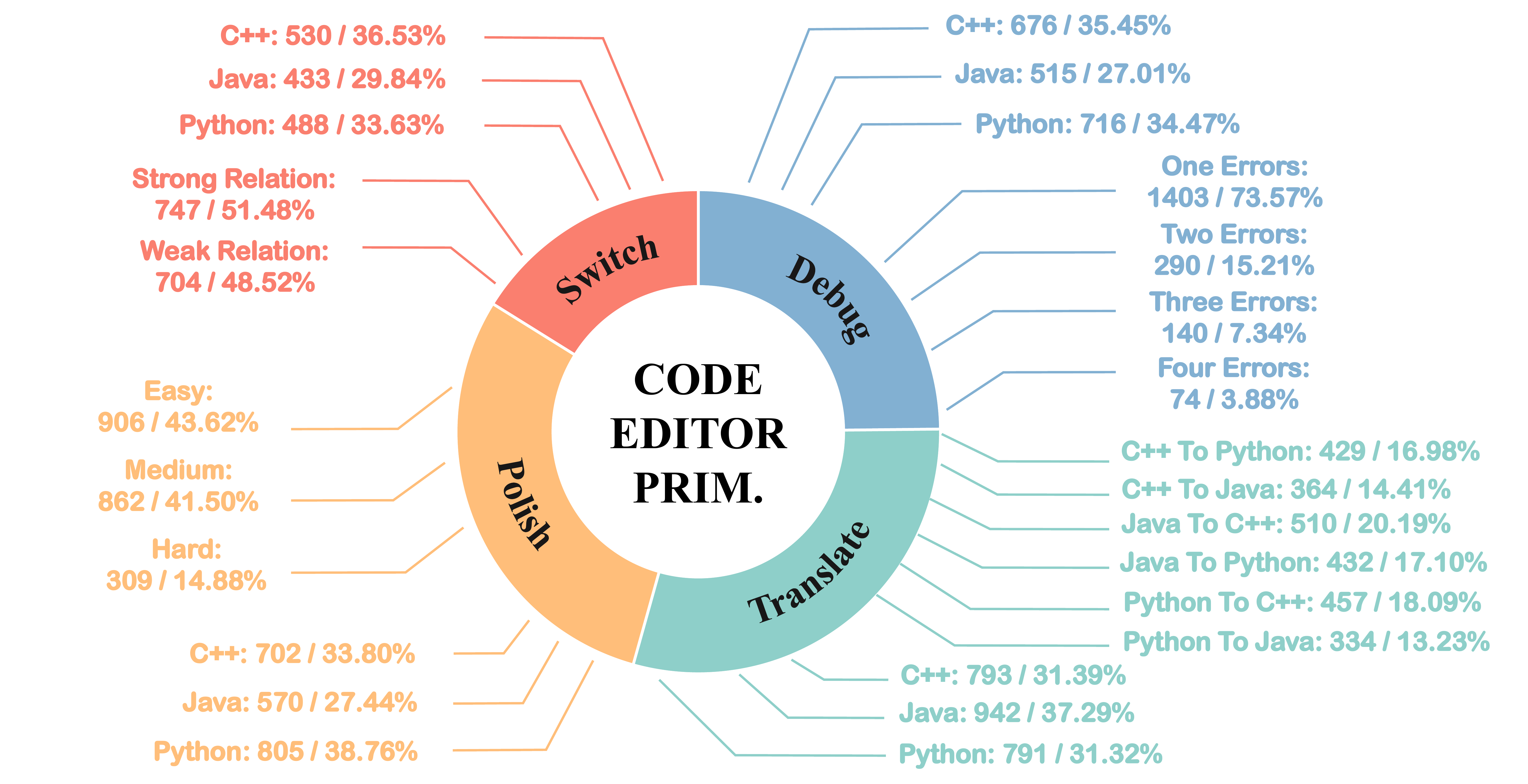}
\end{center}
\caption{Primary Dataset Analysis}
\label{fig:four_code_primary}
\end{figure}

\begin{figure}[h]
\begin{center}
\includegraphics[width=\linewidth]{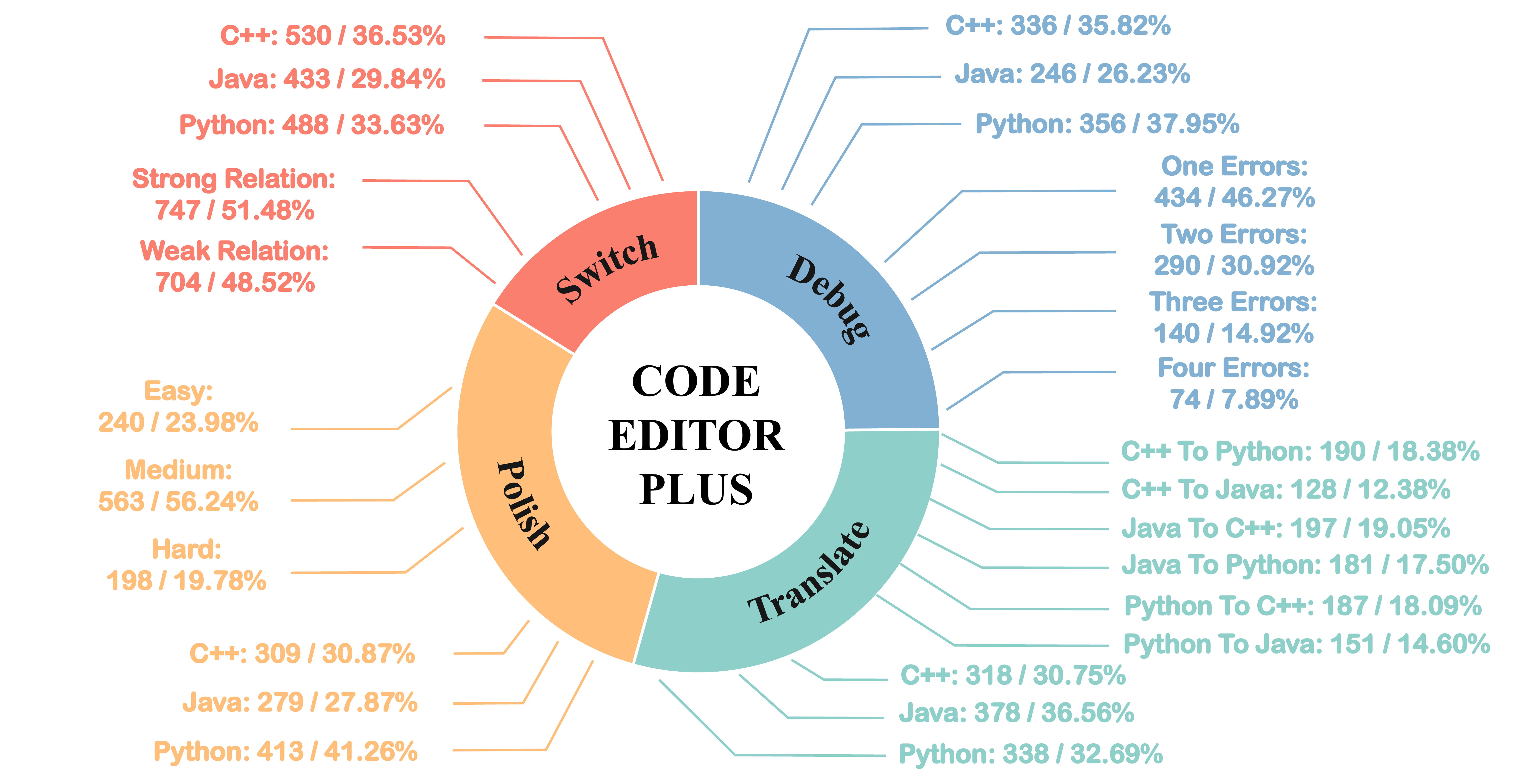}
\end{center}
\caption{Plus Dataset Analysis}
\label{fig:four_code_plus}
\end{figure}

\begin{figure*}[htbp]
    \centering
    \begin{minipage}{0.45\textwidth}
        \centering
        \includegraphics[width=\textwidth]{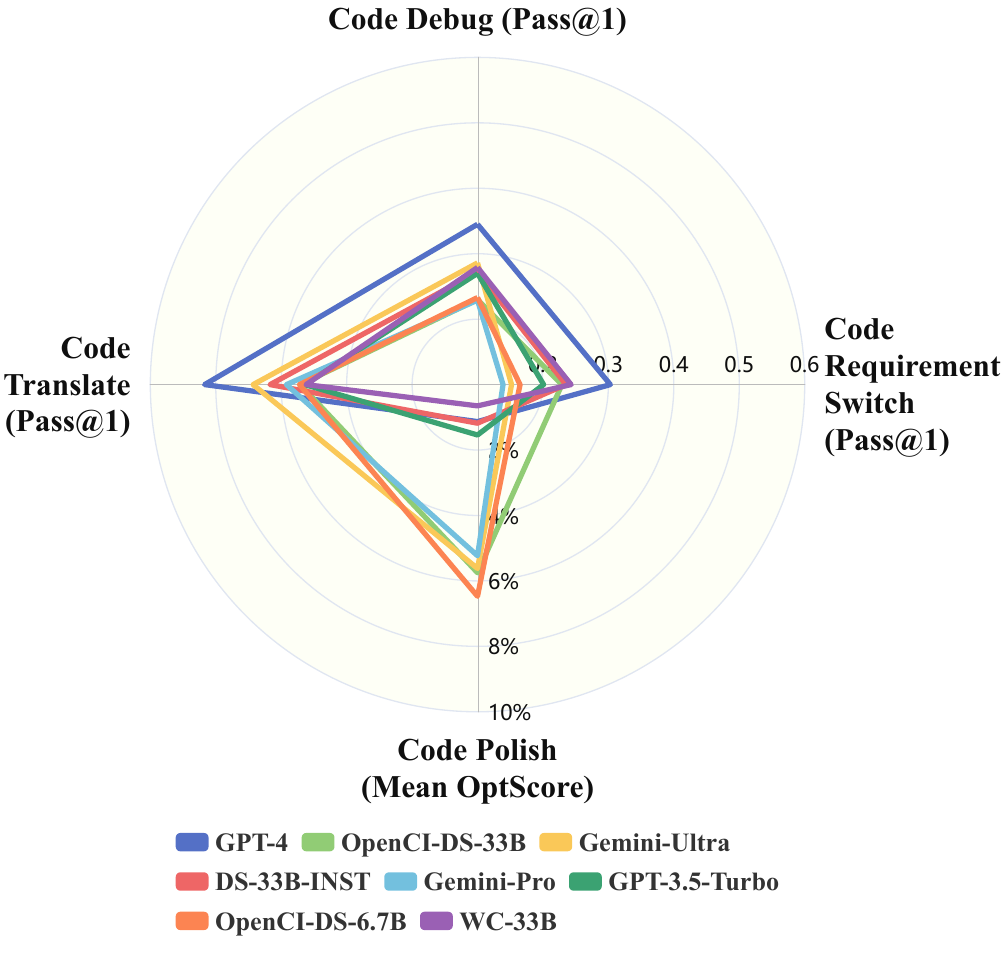}
    \end{minipage}
    \hfill
    \begin{minipage}{0.495\textwidth}
        \centering
        \includegraphics[width=\textwidth]{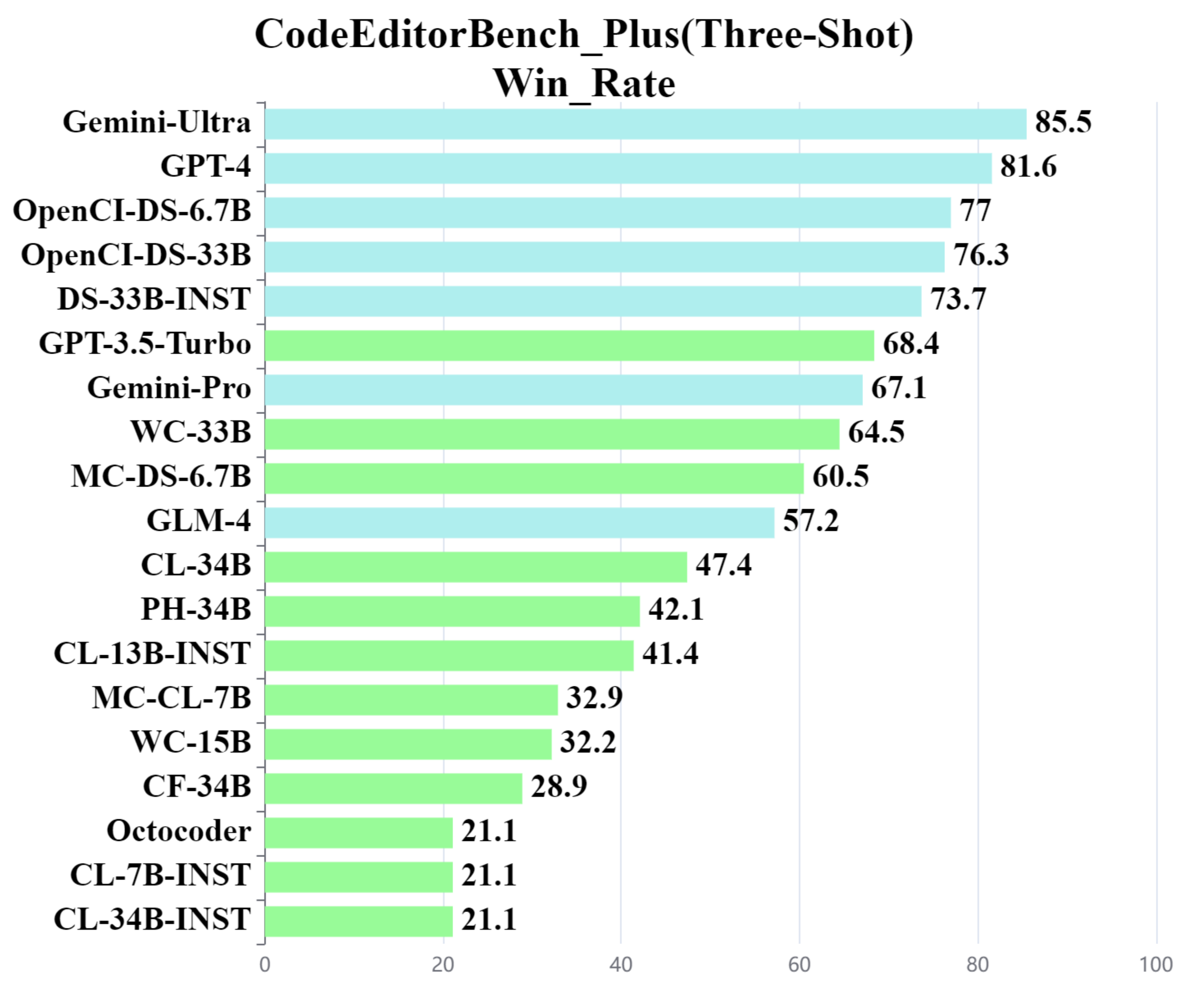}
    \end{minipage}
    \hfill
    \caption{Based on Section \ref{win_rate}. \textbf{Left}. We propose evaluating LLMs across four scenarios capturing various code editing capabilities,namely code debug, code translate, code polish, and code requirement switch. The figure depicts various model performances across the four scenarios available in CodeEditorBench\_Primary in a radial plot – highlighting how relative differences across models change across the scenarios. \textbf{Right}. Performance of open-source and closed-source models on CodeEditorBench\_Primary in three-shot.
    For a comprehensive explanation of the abbreviation, refer to Section \ref{subsec:evaluatedmodels}.}
    \label{fig:mix_few}
    \vspace{-0.2cm}
\end{figure*}

\section{Evaluated Models}
\label{app:Evaluated Models}
We describe the details of models evaluated in our study in Table~\ref{evaluated models}. 
To determine the model release date, we search the model's GitHub repository first. 
If there is no relevant information, the publication date of the paper is used as the release date. 
It is worth noting that for models such as GPT-3.5-Turbo and GPT-4, the referenced date is actually the cut-off date for the model's training data.

\begin{table}[!h]
  \centering
    \begin{tabular}{p{4cm}cccp{4cm}}
    \toprule
    \textbf{Model} & \textbf{Size} & \textbf{Release Date} & \textbf{Open} & \textbf{Link} \\
    \midrule
    ise-uiuc/Magicoder-S-DS-6.7B & 6.7B  & 2023-12-04 & \greencheck     & \textcolor[rgb]{ .02,  .388,  .757}{\href{https://huggingface.co/ise-uiuc/Magicoder-S-DS-6.7B}{Magicoder-S-DS-6.7B}} \\
    ise-uiuc/Magicoder-S-CL-7B & 7B    & 2023-12-04 & \greencheck     & \textcolor[rgb]{ .02,  .388,  .757}{\href{https://huggingface.co/ise-uiuc/Magicoder-S-CL-7B}{Magicoder-S-CL-7B}} \\
    bigcode/octocoder & 15.5B & 2023-08-14 & \greencheck     & \textcolor[rgb]{ .02,  .388,  .757}{\href{https://huggingface.co/bigcode/octocoder}{octocoder}} \\
    WizardLM/WizardCoder-15B-V1.0 & 15B   & 2023-06-16 & \greencheck     & \textcolor[rgb]{ .02,  .388,  .757}{\href{https://huggingface.co/WizardLM/WizardCoder-15B-V1.0}{WizardCoder-15B-V1.0}} \\
    WizardLM/WizardCoder-33B-V1.1 & 33B   & 2024-01-04 & \greencheck     & \textcolor[rgb]{ .02,  .388,  .757}{\href{https://huggingface.co/WizardLM/WizardCoder-33B-V1.1}{WizardCoder-33B-V1.1}} \\
    deepseek-ai/deepseek-coder-33b-instruct & 33B   & 2023-02-01 & \greencheck     & \textcolor[rgb]{ .02,  .388,  .757}{\href{https://huggingface.co/deepseek-ai/deepseek-coder-33b-instruct}{deepseek-coder-33b-instruct}} \\
    codefuse-ai/CodeFuse-CodeLlama-34B & 34B   & 2023-09-11 & \greencheck     & \textcolor[rgb]{ .02,  .388,  .757}{\href{https://huggingface.co/codefuse-ai/CodeFuse-CodeLlama-34B}{CodeFuse-CodeLlama-34B}} \\
    Phind/Phind-CodeLlama-34B-v2 & 34B   & 2023-09-01 & \greencheck     & \textcolor[rgb]{ .02,  .388,  .757}{\href{https://huggingface.co/Phind/Phind-CodeLlama-34B-v2}{Phind-CodeLlama-34B-v2}} \\
    m-a-p/OpenCodeInterpreter-DS-6.7B & 6.7B   & 2024-02-22 & \greencheck     & \textcolor[rgb]{ .02,  .388,  .757}{\href{https://huggingface.co/m-a-p/OpenCodeInterpreter-DS-6.7B}{OpenCodeInterpreter-DS-6.7B}} \\
    m-a-p/OpenCodeInterpreter-DS-33B & 33B   & 2024-02-22 & \greencheck     & \textcolor[rgb]{ .02,  .388,  .757}{\href{https://huggingface.co/m-a-p/OpenCodeInterpreter-DS-33B}{OpenCodeInterpreter-DS-33B}} \\
    codellama/CodeLlama-34b-hf & 34B   & 2023-08-24 & \greencheck     & \textcolor[rgb]{ .02,  .388,  .757}{\href{https://huggingface.co/codellama/CodeLlama-34b-hf}{CodeLlama-34b-hf}} \\
    codellama/CodeLlama-7b-Instruct-hf & 7B    & 2023-08-24 & \greencheck     & \textcolor[rgb]{ .02,  .388,  .757}{\href{https://huggingface.co/codellama/CodeLlama-7b-Instruct-hf}{CodeLlama-7b-Instruct-hf}} \\
    codellama/CodeLlama-13b-Instruct-hf & 13B   & 2023-08-24 & \greencheck     & \textcolor[rgb]{ .02,  .388,  .757}{\href{https://huggingface.co/codellama/CodeLlama-13b-Instruct-hf}{CodeLlama-13b-Instruct-hf}} \\
    codellama/CodeLlama-34b-Instruct-hf & 34B   & 2023-08-24 & \greencheck     & \textcolor[rgb]{ .02,  .388,  .757}{\href{https://huggingface.co/codellama/CodeLlama-34b-Instruct-hf}{CodeLlama-34b-Instruct-hf}} \\
    gpt-3.5-turbo-1106~\citep{openai2022} & -     & 2021-10-01 &   \redmark    & \textcolor[rgb]{ .02,  .388,  .757}{\href{https://platform.openai.com/docs/models/gpt-3-5-turbo}{gpt-3.5-turbo-1106}} \\
    gpt-4-0613~\citep{openai2023} & -     & 2021-10-01 &     \redmark  & \textcolor[rgb]{ .02,  .388,  .757}{\href{https://platform.openai.com/docs/models/gpt-4-and-gpt-4-turbo}{gpt-4-0613}} \\
    glm-4~\citep{zhipuai2024} & -     & 2024-01-16 &  \redmark     & \textcolor[rgb]{ .02,  .388,  .757}{\href{https://open.bigmodel.cn/dev/api}{glm-4}} \\
    gemini-pro~\citep{google2023} & -     & 2023-12-06 &    \redmark   & \textcolor[rgb]{ .02,  .388,  .757}{\href{https://ai.google.dev/models/gemini}{gemini-pro}} \\
    gemini-ultra~\citep{google2023} & -     & 2023-12-06 &  \redmark     & - \\
    \bottomrule
    \end{tabular}
    \caption{\label{evaluated models}Overview of evaluated models.}
\end{table}

\section{Detailed Prompts}
The prompt formats demonstrated here are utilized by closed-source models. The instructions used by open-source models are similar to those of closed-source models, with the main differences being as follows:
\begin{itemize}
    \item Given the limited ability of open-source models to generate code in standard format, we explicitly specify that open-source models generate code enclosed in "\texttt{\textasciigrave\textasciigrave\textasciigrave}", facilitating post processing.
    \item Open-source models typically adhere to a fixed prompt format during the instruction fine-tuning phase, requiring the addition of special identifiers before and after the instruction.
\end{itemize}

\subsection{Code Debug}
\begin{figure}[!h]
\begin{center}
\includegraphics[width=\linewidth]{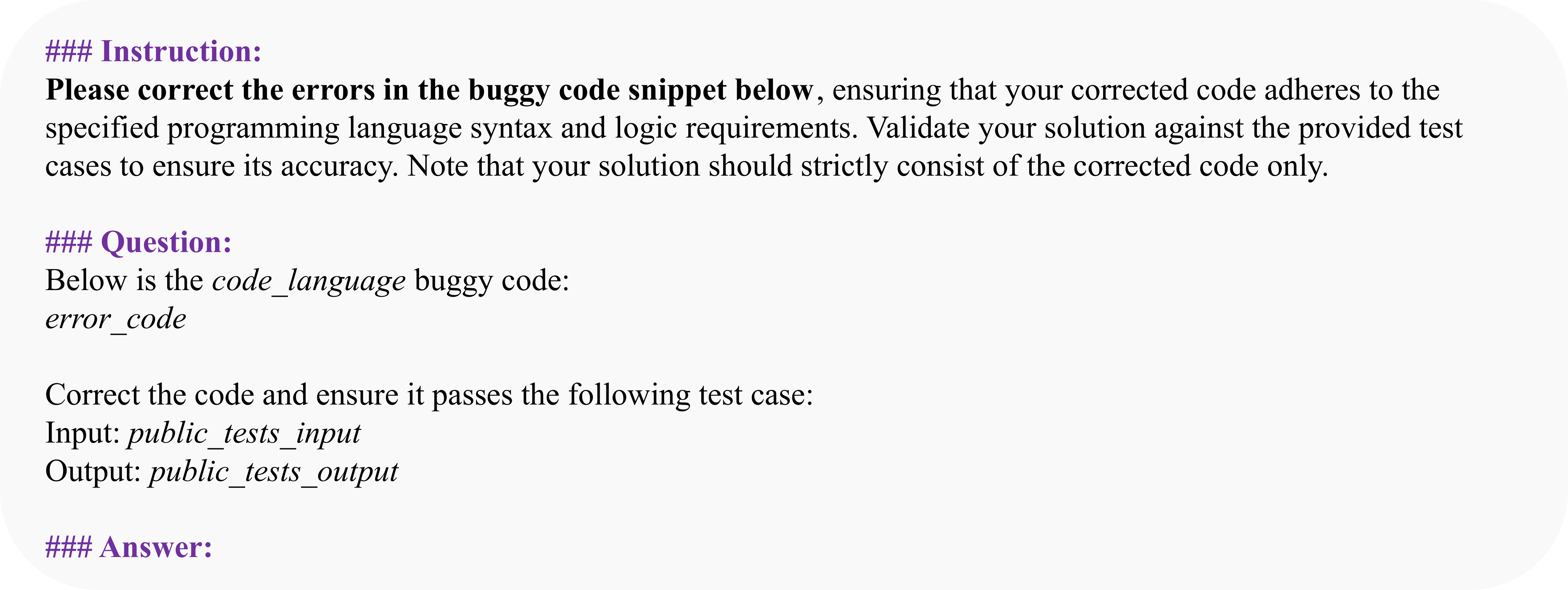}
\end{center}
\caption{Zero-shot Prompt for Closed Models in Code Debug Dataset.}
\label{fig:zero_debug}
\end{figure}

\begin{figure}[h]
\begin{center}
\includegraphics[width=\linewidth]{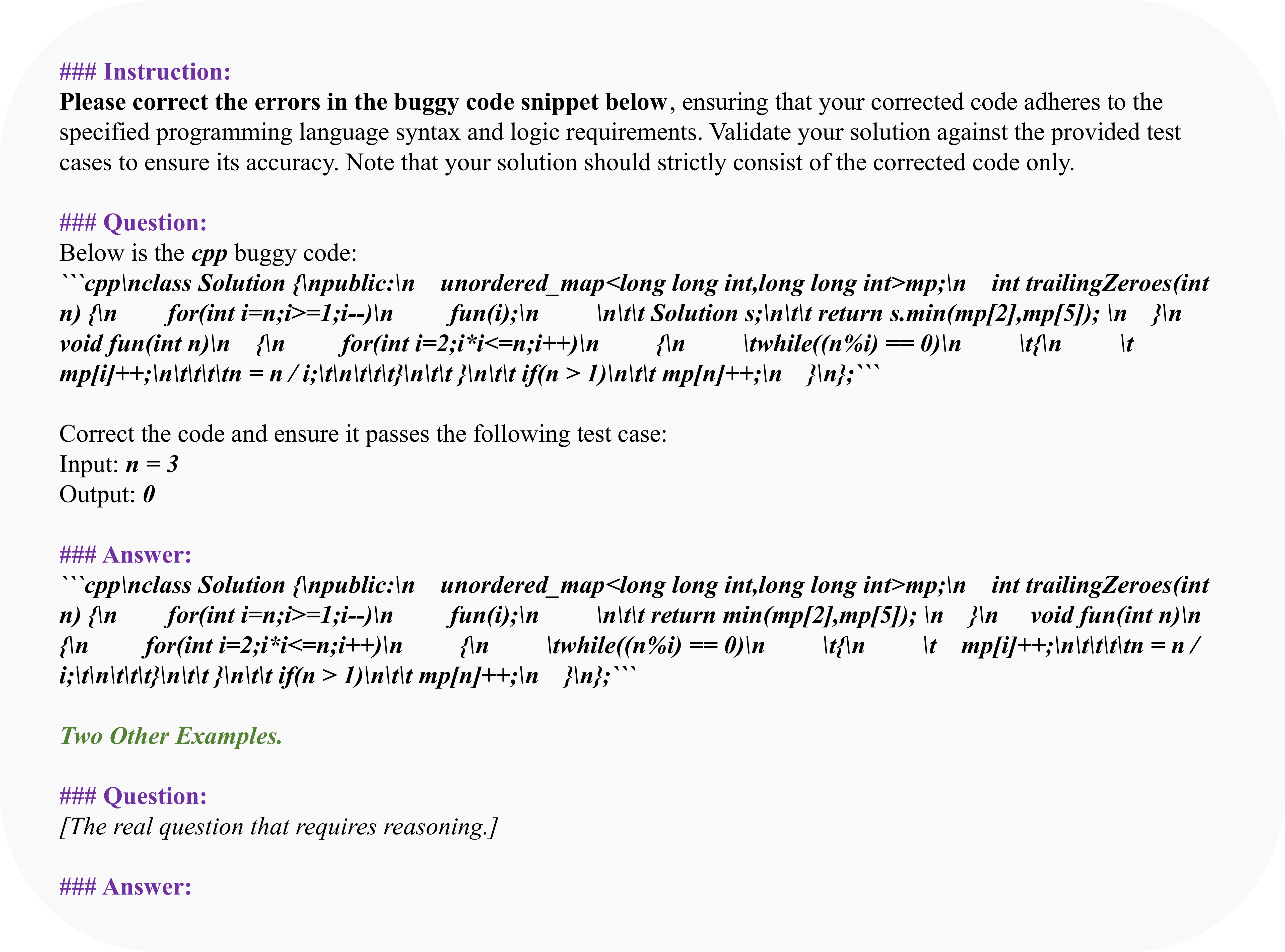}
\end{center}
\caption{Few-shot Prompt for Closed Models in Code Debug Dataset.}
\label{fig:few_debug}
\end{figure}

Below we present the prompt formats used by closed-source models in the code debug scenario, under zero-shot and few-shot.
See details in Figure~\ref{fig:zero_debug},~\ref{fig:few_debug}.

\subsection{Code Translate}
\begin{figure}[h]
\begin{center}
\includegraphics[width=\linewidth]{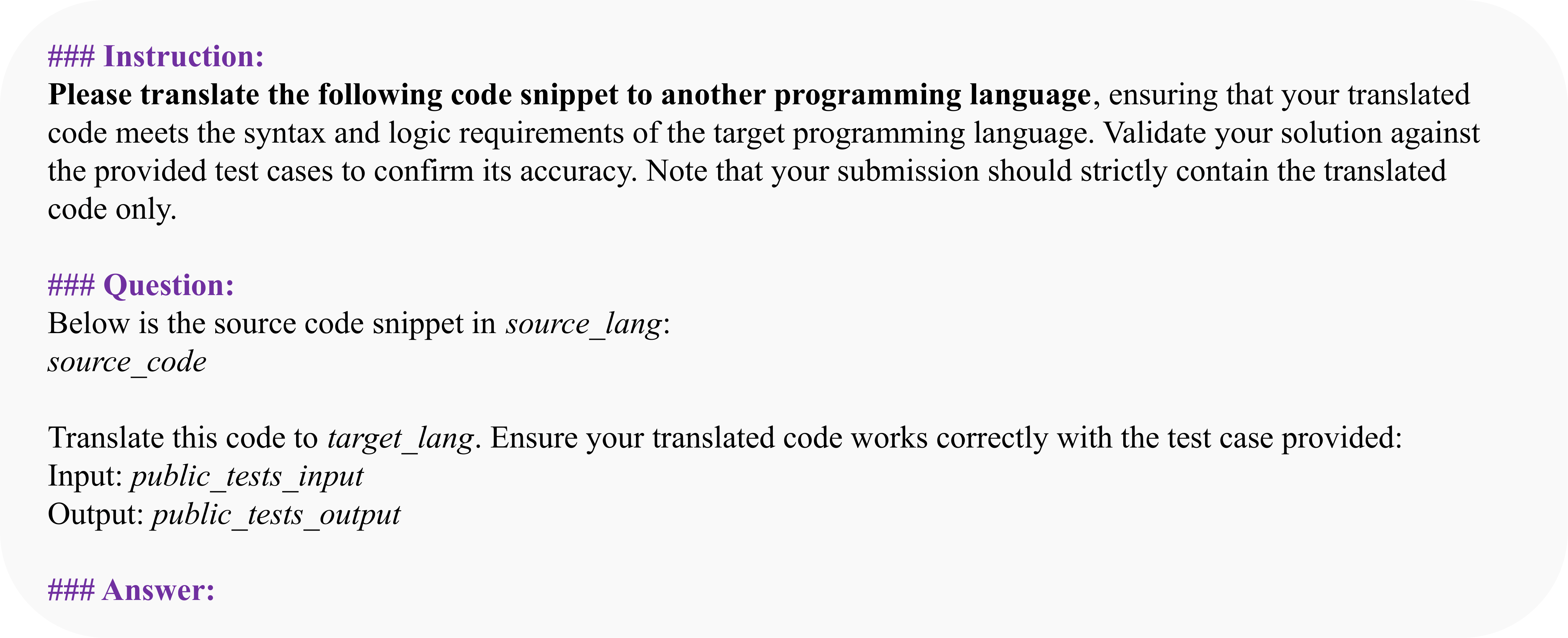}
\end{center}
\caption{Zero-shot Prompt for Closed Models in Code Translate Dataset.}
\label{fig:zero_translate}
\end{figure}

\begin{figure}[h]
\begin{center}
\includegraphics[width=\linewidth]{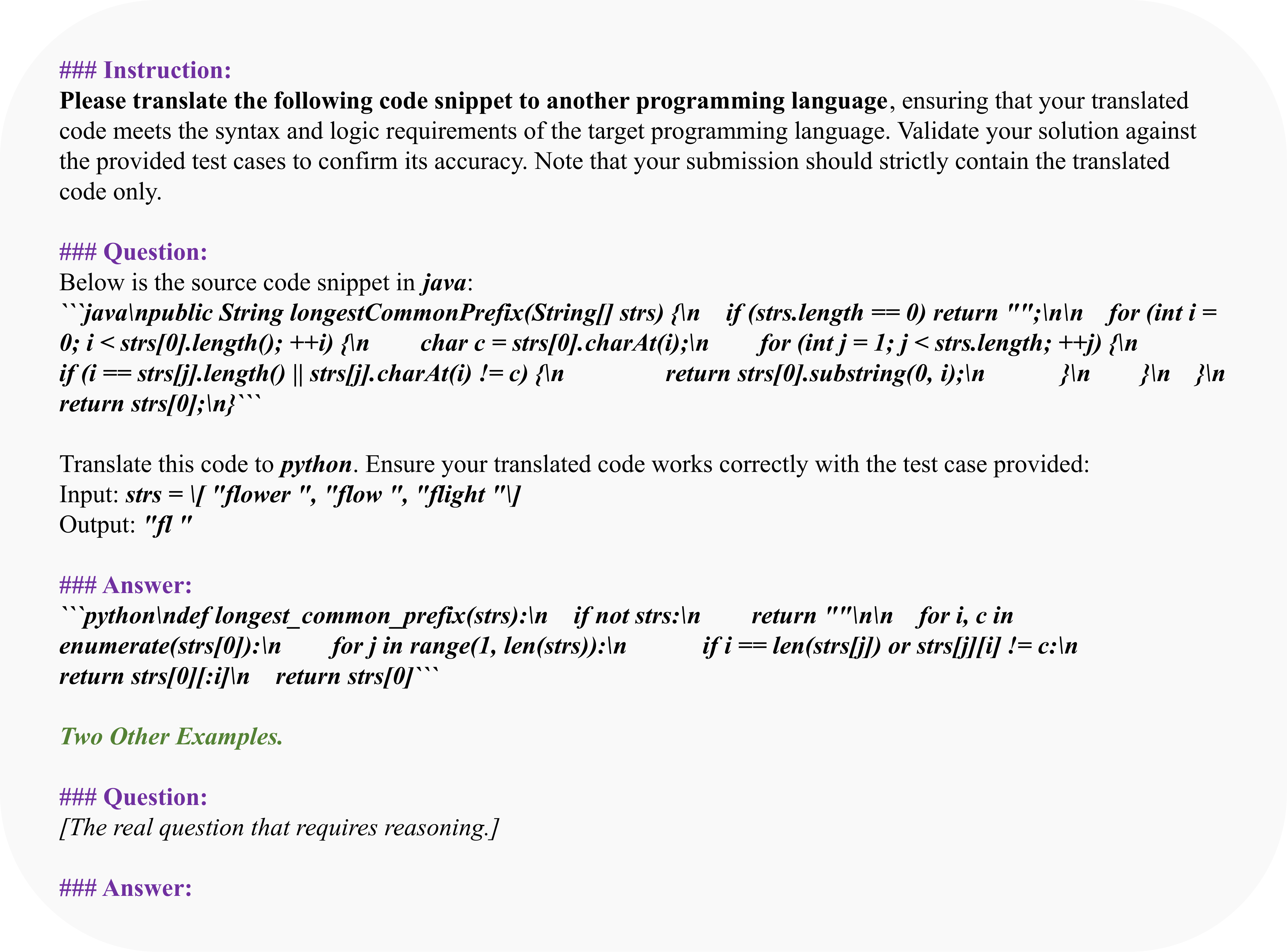}
\end{center}
\caption{Few-shot Prompt for Closed Models in Code Translate Dataset.}
\label{fig:few_translate}
\end{figure}

Below we present the prompt formats used by closed-source models in the code translate scenario, under zero-shot and few-shot.
See details in Figure~\ref{fig:zero_translate},~\ref{fig:few_translate}.

\subsection{Code Polish}
\begin{figure}[h]
\begin{center}
\includegraphics[width=\linewidth]{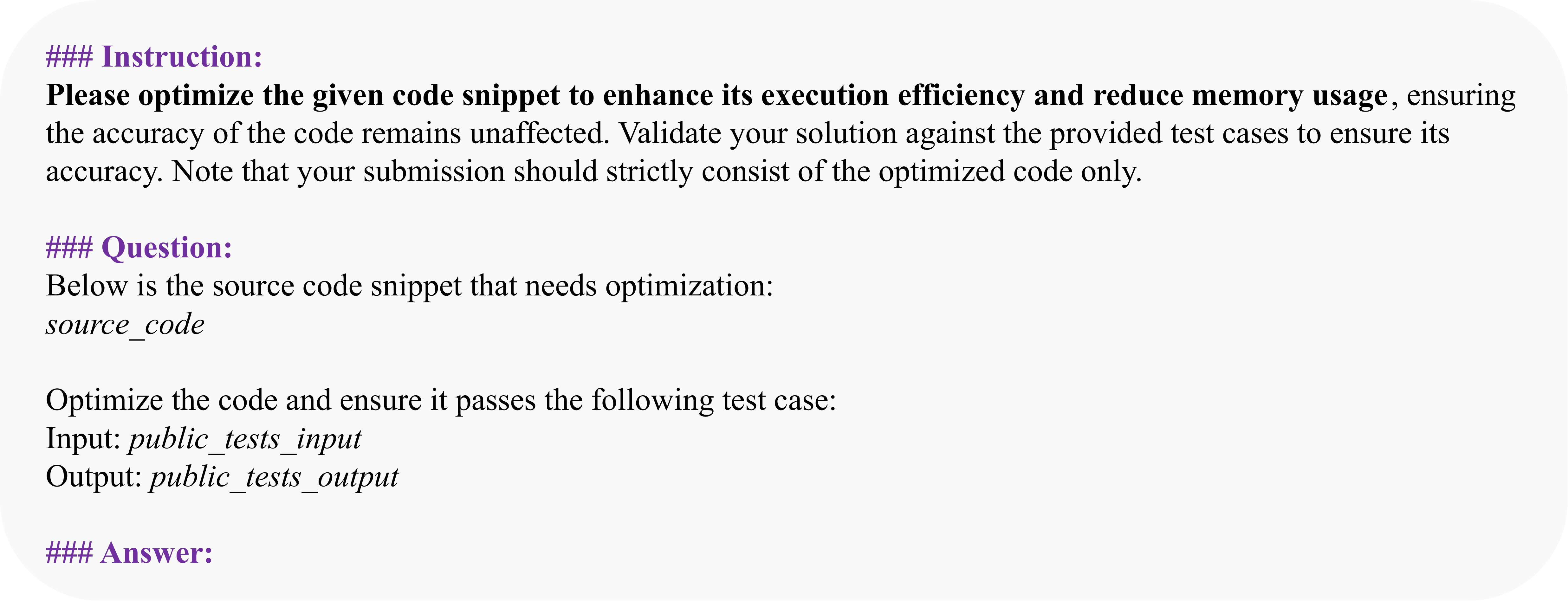}
\end{center}
\caption{Zero-shot Prompt for Closed Models in Code Polish Dataset.}
\label{fig:zero_polish}
\end{figure}

\begin{figure}[h]
\begin{center}
\includegraphics[width=\linewidth]{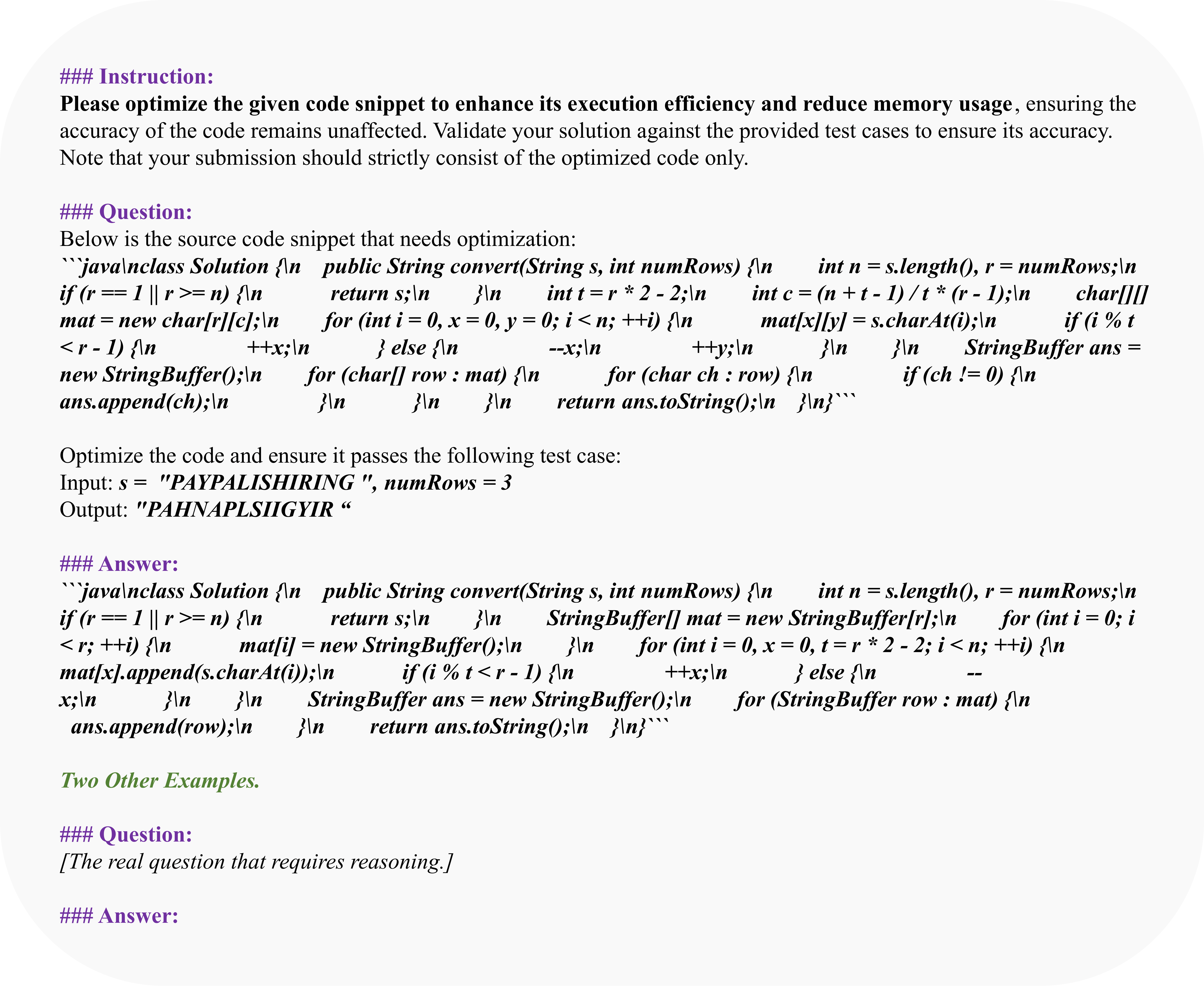}
\end{center}
\caption{Few-shot Prompt for Closed Models in Code Polish Dataset.}
\label{fig:few_polish}
\end{figure}

Below we present the prompt formats used by closed-source models in the code polish scenario, under zero-shot and few-shot.
See details in Figure~\ref{fig:zero_polish},~\ref{fig:few_polish}.

\subsection{Code Requirement Switch}
\begin{figure}[h]
\begin{center}
\includegraphics[width=\linewidth]{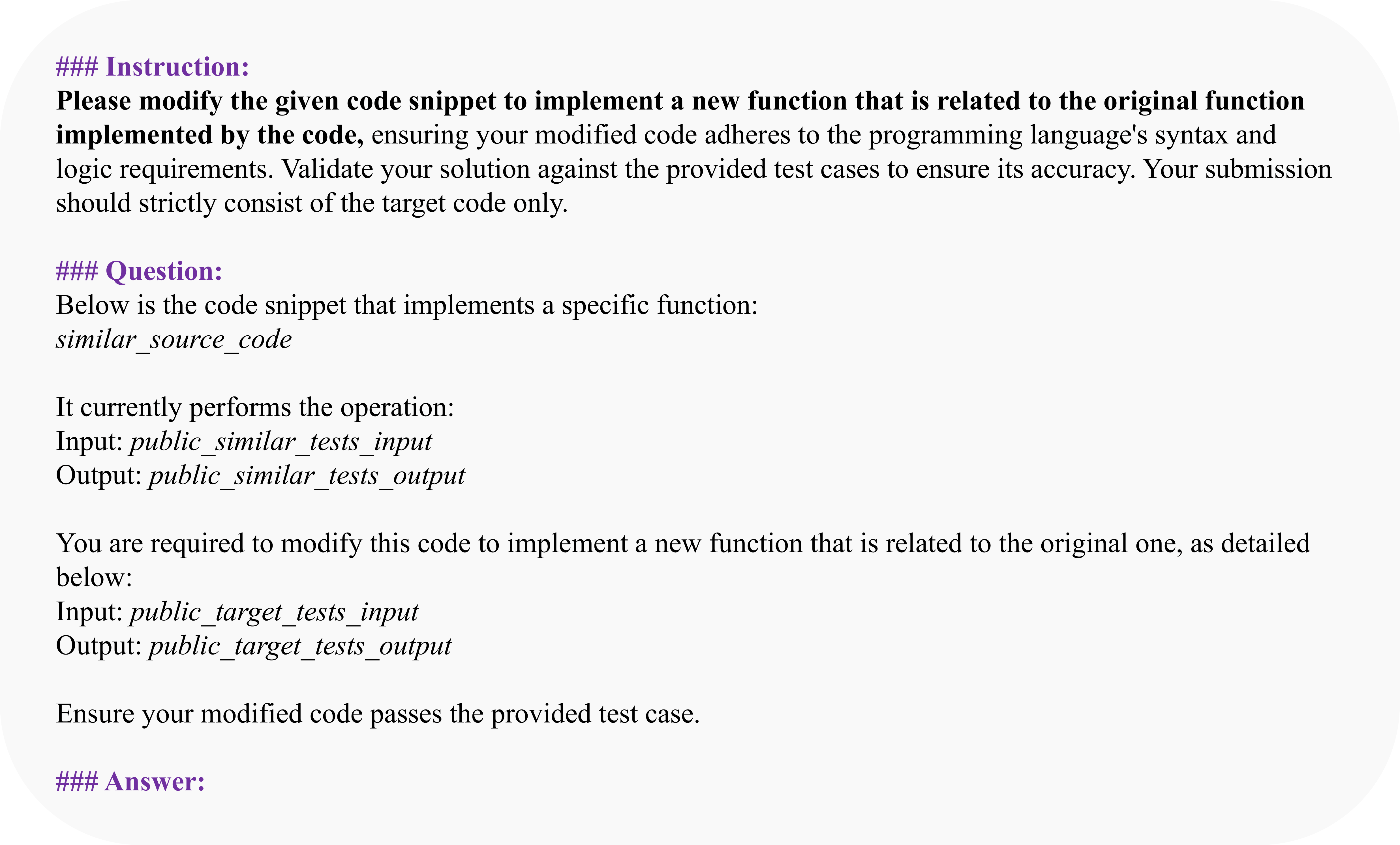}
\end{center}
\caption{Zero-shot Prompt for Closed Models in Code Switch Dataset.}
\label{fig:zero_switch}
\end{figure}

\begin{figure}[h]
\begin{center}
\includegraphics[width=\linewidth]{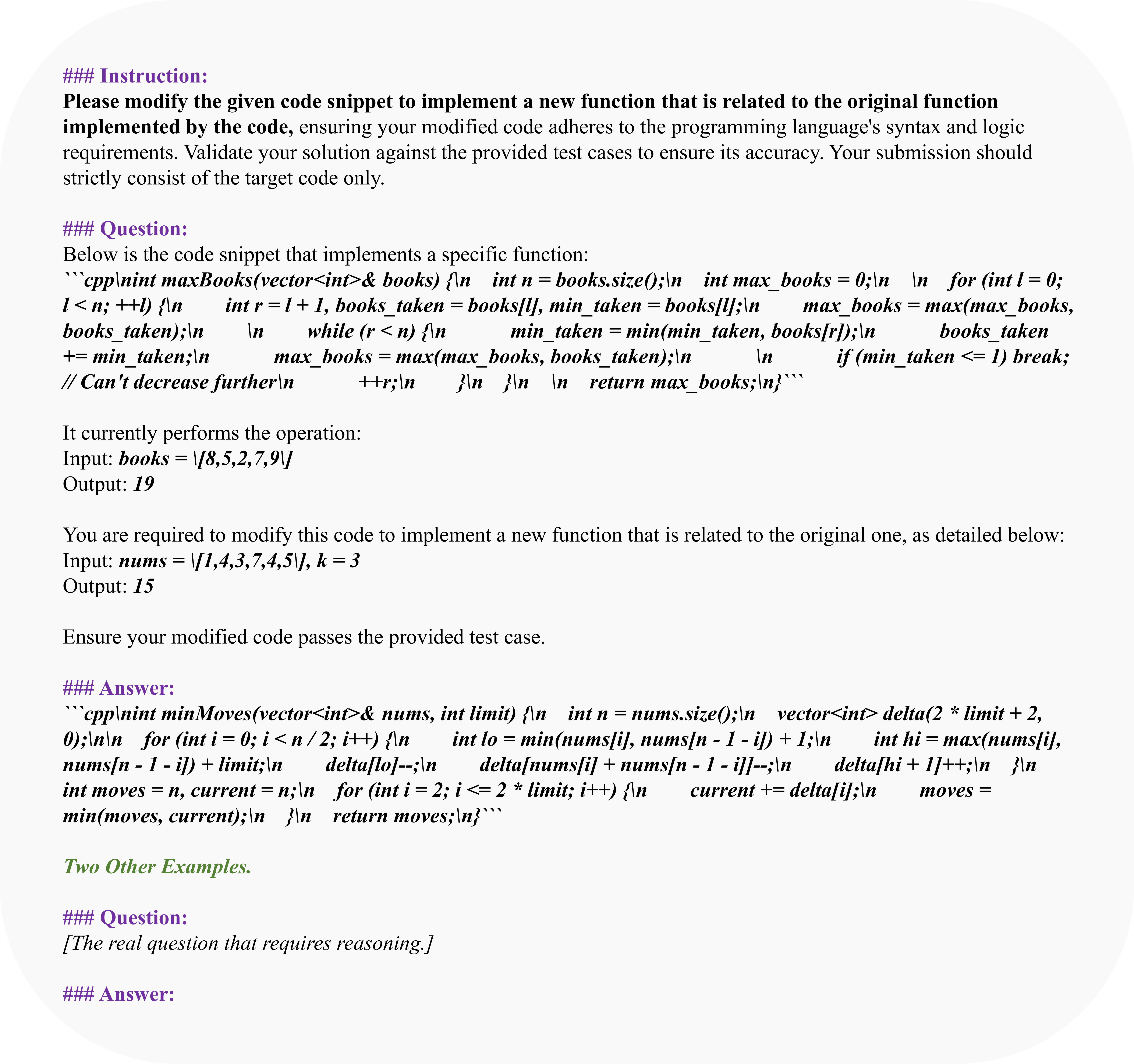}
\end{center}
\caption{Few-shot Prompt for Closed Models in Code Switch Dataset.}
\label{fig:few_switch}
\end{figure}

Below we present the prompt formats used by closed-source models in the code requirment switch scenario, under zero-shot and few-shot.
See details in Figure~\ref{fig:zero_switch},~\ref{fig:few_switch}.

\clearpage
\section{Code Processing and Template Integration}
\label{code_template}
\begin{figure}[h]
\begin{center}
\includegraphics[width=\linewidth]{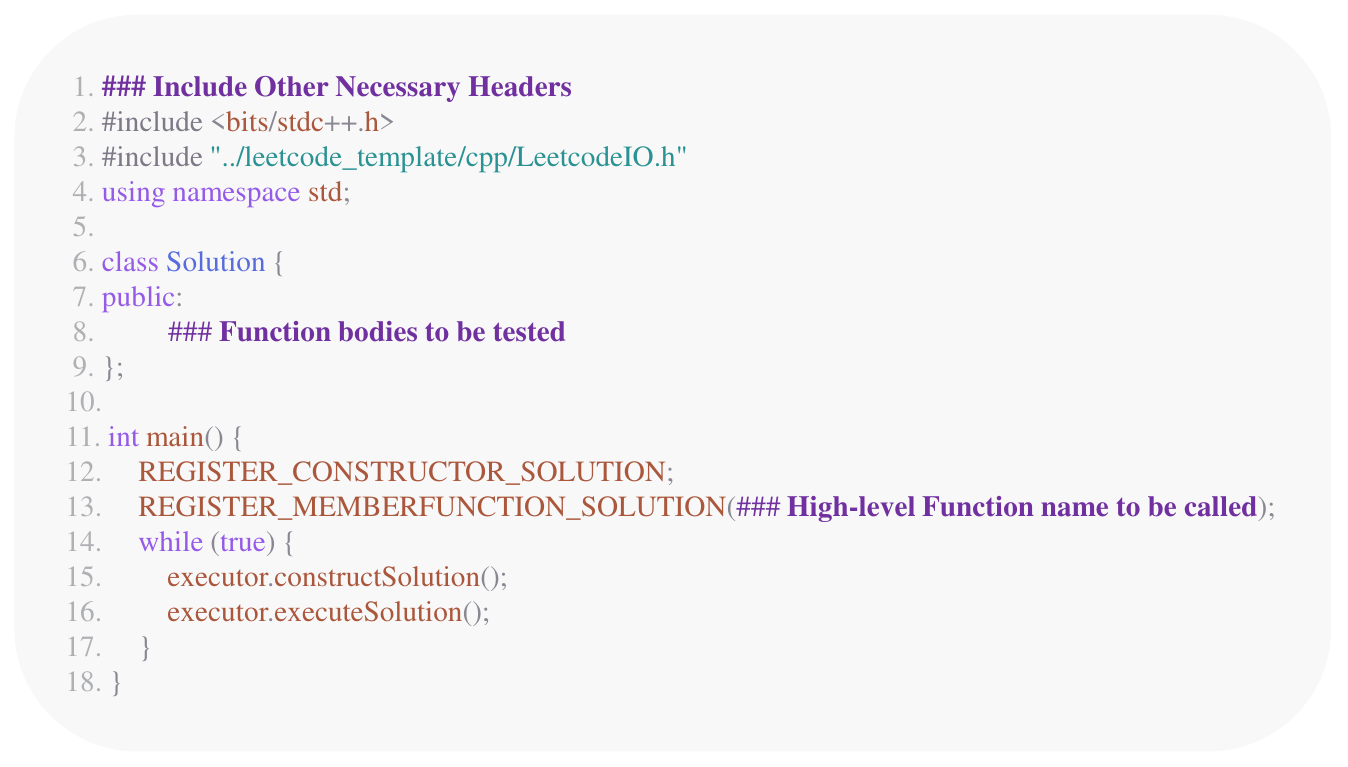}
\end{center}
\caption{C++ Template.}
\label{tab:c++_template}
\end{figure}

\begin{figure}[h]
\begin{center}
\includegraphics[width=\linewidth]{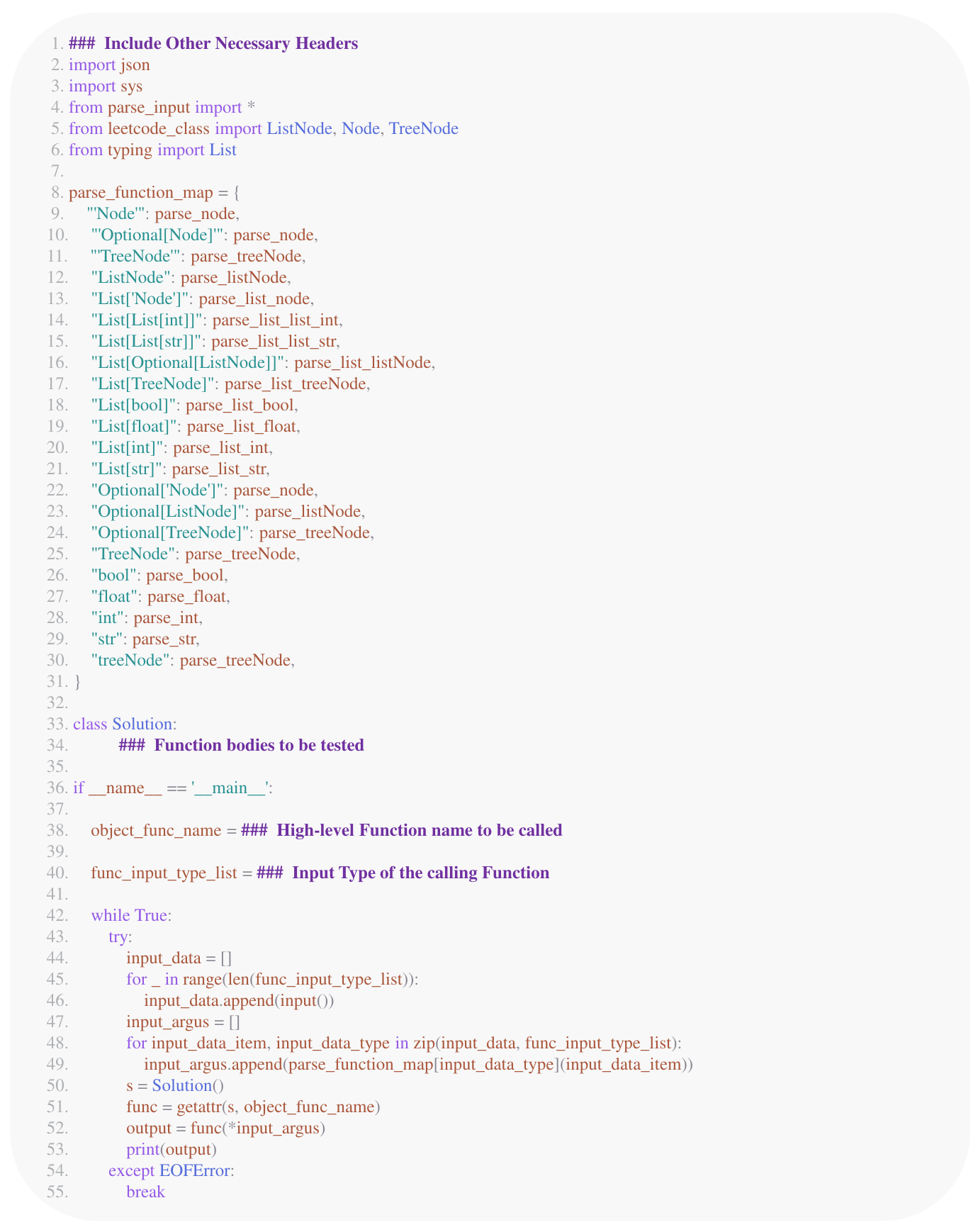}
\end{center}
\caption{Python Template.}
\label{tab:python_template}
\end{figure}

\begin{figure}[h]
\begin{center}
\includegraphics[width=\linewidth]{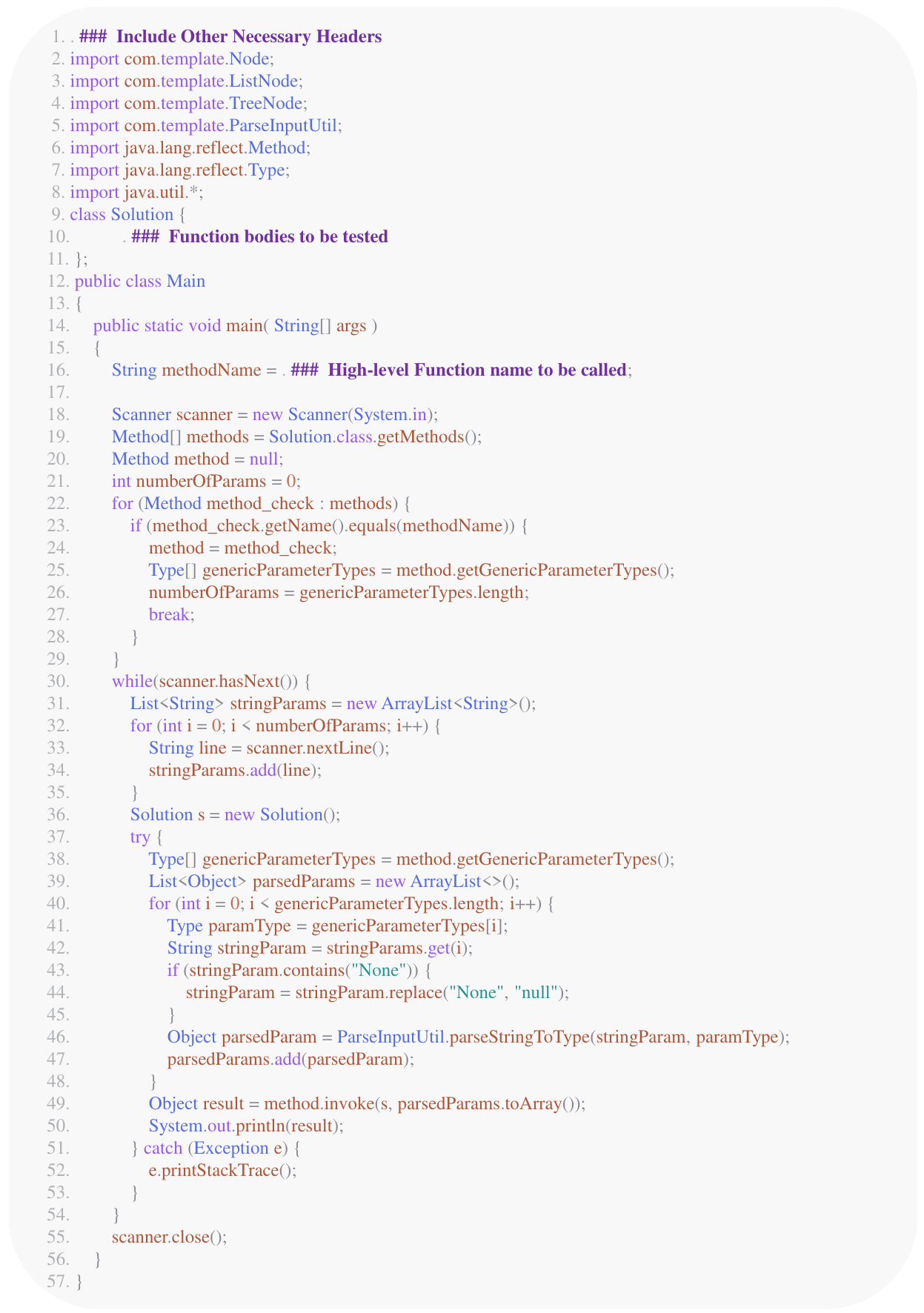}
\end{center}
\caption{Java Template.}
\label{tab:java_template}
\end{figure}

We developed executable templates for C++, Python, and Java, as shown in Figure~\ref{tab:c++_template},~\ref{tab:python_template} and~\ref{tab:java_template}. Submitted code only needs to contain core functions and the high-level function name used for main function calls, aligning with the submission requirements of LeetCode.
For code parsing and function name extraction, all code uses tree-sitter\footnote{\url{https://github.com/tree-sitter/tree-sitter}} - an incremental parsing system for programming tools to retrieve function calls to ensure that the correct high-level function name is obtained.

\clearpage
\section{Pass Criteria}
\label{app:pass}
Table~\ref{tab:pass_criteria} shows our detailed pass criteria.
\begin{table}[htbp]
\centering
\renewcommand{\arraystretch}{2}
\scalebox{0.8}{
\begin{tabular}{@{}lccc@{}}
\hline
\textbf{Scenario} & \textbf{Time limit} & \textbf{Memory limit} & \textbf{Pass Criteria} \\
\hline
Debug & \multirow{3}{*}{\parbox[c]{3.5cm}{%
    300s for all test cases.\\
    30s for single test case.%
}} & \multirow{3}{*}{$512$MB} & Pass all test cases. \\


{Translate} & & & {
Pass all test cases in target language.}\\


Switch & & & Pass all test cases. \\
\hline
\\[0.5ex]
\multirow{3}{*}{Polish} & \multirow{3}{*}{$\bar{T}$ms} & \multirow{3}{*}{$\bar{M}$MB} &
\parbox{7cm}{\centering Pass all test cases \\[0.5ex]
                and \\[0.5ex]
                $\bar{T}_{\text{avg}} < \bar{T}$ or $\bar{M}_{\text{avg}} < \bar{M}$} \\[0.5ex]
 & & \\[0ex]
 & & \\[0ex]
\hline
\end{tabular}
}

\caption{Pass criteria in four scenarios. $\bar{T}$ and $\bar{M}$ are obtained by averaging 20 runs of the standard code. During the judging process, the LLM-generated code is run twice to obtain $\bar{T}_{\text{avg}}$ and $\bar{M}_{\text{avg}}$.}
\label{tab:pass_criteria}
\end{table}

\section{Evaluation Configuration}
\label{app:eval_configure} 
Our OJ is built on a server equipped with a high-performance Intel(R) Xeon(R) Platinum 8480C processor boasting 224 cores. 
It supports up to 4 petabytes (PB) of physical memory and has a virtual memory space of up to 128 terabytes (TB). 
The judging environment for C++ is based on the g++ compiler, version 9.4.0, utilizing the C++17 standard. 
For Python, the judging environment is based on Python 3.8.10, and for Java, it relies on OpenJDK version 11.0.22. 
The previously mentioned template utilizes the gson-2.9.1.jar library to process the input.

\end{document}